# Fluorescent nano- and microparticles for sensing cellular microenvironment: past, present and future applications

Giuliana Grasso,*[a] Francesco Colella,[a] Stefania Forciniti,[a] Valentina Onesto,[a] Helena Iuele,[a] Anna Chiara Siciliano,[a,b] Federica Carnevali,[a,b] Anil Chandra,[c] Giuseppe Gigli,[a,b] and Loretta L. del Mercato*[a]

The tumor microenvironment (TME) features distinct hallmarks, including acidosis, hypoxia, reactive oxygen species (ROS) generation, and altered ion fluxes, which are crucial targets for early cancer biomarker detection, tumor diagnosis, and therapeutic strategies. A variety of imaging and sensing techniques have been developed and employed in both research and clinical settings to visualize and monitor cellular and TME dynamics. Among these, ratiometric fluorescence-based sensors have emerged as powerful analytical tools, providing precise and sensitive insights into the TME and enabling real-time detection and tracking of dynamic changes. In this comprehensive review, we discuss the latest advancements in ratiometric fluorescent probes designed for optical mapping of pH, oxygen, ROS, ions, and biomarkers within the TME. We elucidate their structural designs and sensing mechanisms, as well as their applications in *in vitro* and *in vivo* detection. Furthermore, we explore integrated sensing platforms that reveal the spatiotemporal behavior of complex tumor cultures, highlighting the potential of high-resolution imaging techniques combined with computational methods. This review aims to provide a solid foundation for understanding the current state of the art and the future potential of fluorescent nano- and microparticles in the field of cellular microenvironment sensing.

## 1. Introduction

Cancer is one of the leading causes of death globally and has become more widespread over last several decades due to unhealthy lifestyle and environmental factors causing more mutations to happen, increasing likelihood of cancer development.[1] According to statistical studies recently published by the American Cancer Society, 16.2 million of cancer-related deaths and 28 million of new cancer cases are projected to be the global burden by 2040.[2,3] Moreover, an event that worsened cancer incidence in the last three years has been the coronavirus disease 2019 (COVID-19) pandemic, which made register an uptick in advanced stage of tumor cases, accomplished with increased mortality.[4]

One of the major steps to prevent cancer associated mortality and increasing the chances of successful treatment is early detection of tumors to this end, the gold standard methods employed to detect a tumor site are, mainly based on medical imaging techniques, where the most common methods are magnetic resonance imaging (MRI)[5], positron emission tomography (PET)[6] coupled to computed tomography (CT),[7] ultrasound scanning[8] and photoacoustic endoscopy.[9] Nevertheless, these techniques present some limitations, such as are poor compliance from the patient's point of view, lack of long-term stability of contrast agents and tracers, and requirement of highly qualified personnel. Together with cancer oncologists, scientists are directing significant resources and efforts towards the study of biological features of the tumors that can promote early detection. In this context, the complex tumor microenvironment (TME), in which cancer cells and rich stroma interact with each other releasing growth factors, proteins and membrane-derived vesicles, provides new avenues for early diagnosis.[10–15] Different strategies targeting the TME have emerged since its importance in influencing therapeutic outcome.[16,17] Notably, the TME is highly heterogeneous and dynamic, and it is characterized by the establishment of pH and oxygen gradients resulting from an increased cellular metabolic activity and from an altered blood perfusion. This marked heterogeneity significantly affects the efficacy of anticancer treatments.[18] Dynamic mapping of TME's parameters, such as pH and oxygen, is crucial for understanding their role in cellular and subcellular processes because it could help to better understand the link between pH/oxygen distribution, cell morphology and cell functions. In this context, numerous sensing systems have been developed to allow prompt detection of key analytes in the TME, in order to obtain a valid metabolic read out for cancer diagnosis and treatment.[19,20] Among these, ratiometric fluorescence-based nano- and micro-sensors stand out as valid and non-invasive approaches for characterizing cellular microenvironments and sub-cellular compartments with high precision over time and space. These technologies pave the way for the powerful era of modern precision and personalized medicine.

*Deep insight in TME*

The genesis and the development of cancer disease involves multistep processes that start with genetic or epigenetic changes in tumor cells,[21–23] followed by a dynamic crosstalk leading to the rearrangement of a tumor-supportive and highly reactive microenvironment (TME) that surrounds the tumor (**Figure 1**).[24–26] It is now widely recognized that TME plays a crucial role in cancer initiation,[27] progression[28] and metastasis.[29]

The main constituents of the TME are cancer cells[30,31] and accessory cells, that include cancer associated fibroblasts (CAFs),[32,33] immune and inflammatory cells,[34,35] all embedded in a dense stroma of extracellular matrix (ECM) components such as collagen type I,[36,37] fibronectin,[38] hyaluronic acid[39] and growth factors.[40] ECM not only functions as a support for tumor cells but also regulates and promotes cell-cell and cell-matrix interactions.[41] Additionally, the ECM is involved in signalling pathways that regulate cell behaviour and differentiation, so changes in the ECM can disrupt normal cellular processes and lead to disease development and progression.[42] Together, the interplay between cancer cells and ECM contributes to the rise of the tumor heterogeneity, which is considered the major cause of treatment failure in current therapies.[43] This feature is not only a consequence of clonal outgrowth of cells with genetic alterations, but also of epigenetic alterations promoted by several physical and biochemical signals from the TME.[44] The unlimited multiplication of tumor cells is a phenomenon strictly related to their ability to elude growth suppressors[45] and apoptotic signals.[46] As a result, the TME promote a sequence of physical (acidosis, hypoxia, temperature and stiffness)[47,48] and biochemical (adhesion proteins, glycoproteins and proteoglycan, secreted factors, growth factors, and matrix degradation enzymes)[49] adaptations which promote angiogenesis,[50] invasion[51] and metastasis.[52] Moreover, to survive in a hostile microenvironment, which is characterized by high deprivation of oxygen and nutrients, and to maintain an high proliferative rate, some tumor cells are known to adjust their metabolism,[53,54] from the oxidative phosphorylation towards the aerobic glycolysis, the so-called "Warburg effect",[55] which was first observed by Otto H. Warburg in the early twentieth century.[56] In normal cells with adequate oxygen levels, the pyruvate produced by the breakdown of glucose during the glycolysis process, could enter into the tricarboxylic acid (TCA) cycle to generate energy.[57] Tumor cells instead exhibit increased glycolysis activity regardless of the amount of oxygen and produce lactate by activating lactate dehydrogenase and inhibiting mitochondrial metabolism.[55] The resulting acidosis effect is a direct consequence of the lowering of extracellular pH from physiological pH 7.4 to values up to 5.0, whereas the intracellular pH is increased compared to normal cells. As a consequence, the acidity of the interstitial space and the high intracellular pH affect the dynamic and functional cell-cell or cell-matrix crosstalk.[58] The low extracellular pH is an important factor for inducing more aggressive cancer phenotypes, increasing cell motility, extracellular matrix degradation and modifying cellular and intercellular signaling.[59,60] Furthermore, the accumulation of protons ($H^+$) in the extracellular environment is spatially and temporally heterogeneous and influences the efficacy of chemotherapeutic treatments.[61] Additionally, it is well established that also intracellular pH is dysregulated in cancer.[62,63] Although many biological mechanisms contribute to intracellular pH dynamics, the main regulators are the plasma membrane ion exchangers, such as $Na^+/H^+$ exchanger 1 (NHE1), and plasma membrane ion transport proteins including V-ATPases and the monocarboxylate transporters (MCTs). Changes in their expression and activity to facilitate $H^+$ efflux contribute to maintain the alkaline intracellular pH and the acidic extracellular pH in tumor cells.[64]

Furthermore, the higher intracellular pH promotes many cancers behaviors such as increased proliferation, migration, epithelial plasticity and the oncogenic and tumor suppressor functions of mutated proteins.[65] Notably, in solid tumor the low vascularization due to impaired vascular network with formation of abnormal blood vessels and the reduced perfusion of oxygen within the TME equally contribute to the metabolic switch of the cancer cells, promoting the acidification of the TME.[66] This phenomenon, known as *hypoxia*, is mainly induced by hypoxia-inducible factors (HIFs), recognized as master regulator of cell metabolism.[67] The reduction of oxygen levels in the TME is associated with angiogenesis activation and increasing tumor survival, invasiveness, metastatic potential and hamper the therapeutic response.[68]

The hypoxic microenvironment induces also the expression of genes that sustain tumor progression and the generation of reactive oxygen species (ROS).[69] '*Reactive oxygen species*' term groups together two classes of molecular oxygen products derived from reduction-oxidation reactions or electronic excitation during aerobic cellular respiration, which are namely non-radical and free radical species. Examples of ROS are hydrogen peroxide ($H_2O_2$), its reduction product hydroxyl radical (•OH), and superoxide anion radical ($O_2^{-}$).[70–72] Such side products derive from the cellular respiration, carried out in mitochondria, and in physiological conditions are kept under control by detoxification mechanisms.[73,74] The maintenance of their concentration within physiological levels ($10^{-8}$ M for $H_2O_2$ and $10^{-11}$ M for $O_2^{-}$) is called *oxidative eustress*.[75] In this phenomenon, more than 40 enzymes, in particular NADPH oxidase enzymes, are involved in the redox signalling pathways that promote proliferation, differentiation, migration and angiogenesis.[76] In contrast, elevated levels of ROS in the cellular microenvironment determine *oxidative distress*, a condition in which unspecific proteins oxidation leads to reversible or irreversible damages of biomolecules causing pathological states that include inflammation, tumor growth, metastasis and cell death.[77] In the cancer field, the study of ROS has attracted more and more interests in last twenty years, since it is now generally recognized that the regulation of oxidative stress represents a key factor of tumor development and its responses to anticancer therapies. Gorrini's group[78] remarked that moderate concentrations of ROS may contribute to tumor progression since they act as signalling molecules and promote the mutation of nuclear DNA and mitochondrial DNA (mt-DNA). Moreover, concomitant conditions in the TME, such as hypoxia, metabolic defects inducing the Warburg effect,[79] endoplasmic reticulum (ER) stress, and activation of oncogenes, cooperate to the production and dramatic increase of ROS concentration.

In healthy cells, inorganic ions can play different roles in the homeostasis of human body. For instance, inorganic metal cations as magnesium ($Mg^{2+}$), zinc ($Zn^{2+}$), iron ($Fe^{2+}$) and copper ($Cu^{2+}$) are essentials in enzymatic reactions by acting as cofactors, while the cations calcium ($Ca^{2+}$), sodium ($Na^+$) and potassium ($K^+$), and the anion chloride (Cl$^-$) are involved in the electrophysiological events.[80] During the malignant transformation of healthy cells into cancer cells, genes' mutations affect also those encoding for plasma membrane ion channels, thus resulting in the alteration of the ions fluxes, cell membrane potentials ($V_{mem}$), and consequently modifications in the intracellular signalling pathways.[81–83] The disruption of ion homeostasis, pertaining to the transport of ions through channels and their concentration within the TME, gives rise to biophysical phenomena such as elevated pressure, increased stiffness, and mechanical stress. These alterations subsequently lead to the activation or attenuation of molecular signaling pathways implicated in cancer initiation, promotion, and invasion processes.[84] One indicator of normal cell transitioning into a cancerous and proliferative tissues is the more positively charged or depolarized membrane, with a $V_{mem}$ rising from -60 mV to -10/-30 mV in more undifferentiated cancer stem-cells.[85,86] The cytosolic Ca$^{2+}$ levels are important for the integrin-signalling pathway, which is activated to allow cell-cell and cell-ECM communication. The cytosolic Ca$^{2+}$ levels are important for the integrin-signalling pathway, which is activated to allow cell-cell and cell-ECM communication. Dysregulation of calcium ions within cells contributes to cancer-related processes. Specifically, intracellular Ca$^{2+}$ concentration plays a crucial role in regulating cytoskeletal dynamics, which are involved in extracellular matrix (ECM) degradation and the initiation of the metastatic process. This occurs through the activation of epithelial-mesenchymal transition (EMT) pathways and the enzymatic activity of metalloproteases.[87,88] Strictly linked to Ca$^{2+}$ functions are the fluxes of Na$^+$ cations, which are involved in the synergistic activity of Na$^+$/Ca$^{2+}$ exchangers.[89] Furthermore, alterations in intracellular Na$^+$ concentrations lead to a reduction of H$^+$ in the vicinity of cancer cells, resulting in the formation of integrin-mediated focal adhesion contacts that promote cell adhesion.[90] Concomitant with alterations in the membrane potential of cancer cells, K$^+$ fluxes are linked to proliferation, as fluctuations in K$^+$ levels interact with extracellular signal-regulated kinase (ERK 1/2) and c-Jun N-terminal kinase (JNK) signalling pathways.[91,92] Additionally, the Cl$^-$ anion, which typically facilitates the transport of cations such as Ca$^{2+}$, Na$^+$, and K$^+$, can promote migration and metastasis by modulating cell volume.[93]

*Fluorescent ratiometric sensors for TME investigation and mapping*

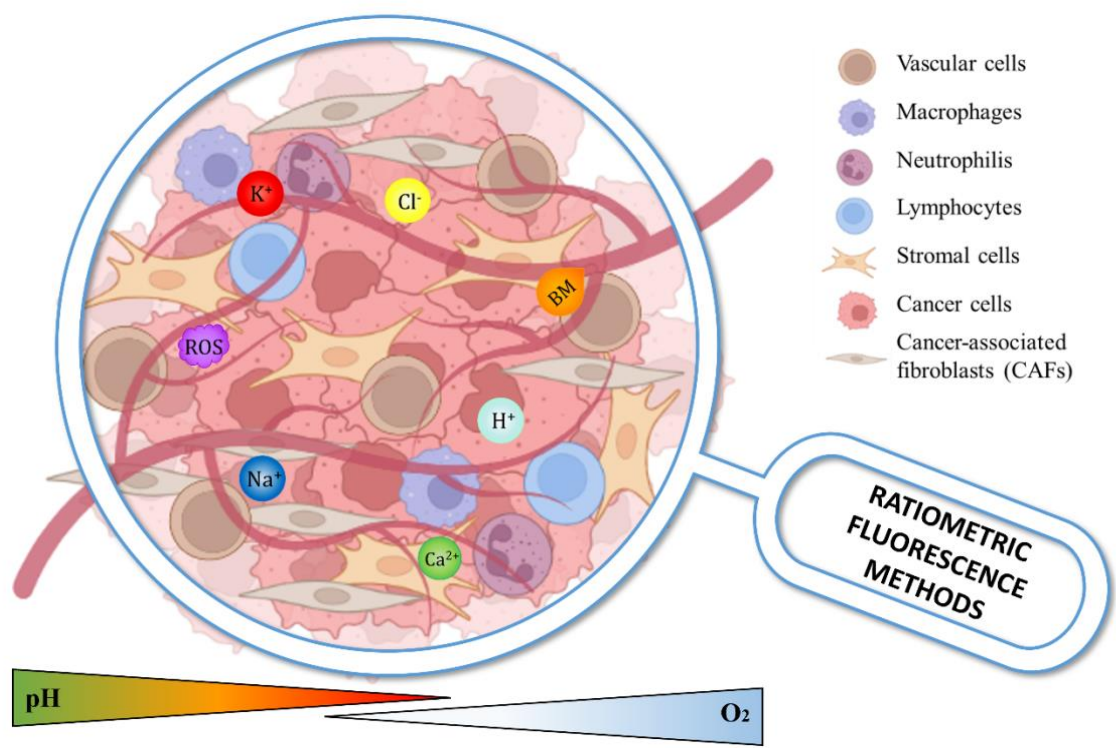

**Fig. 1** A close-up look at the different characteristics and analytes having a biological significance within the TME, which can be precisely examined by ratiometric fluorescence methods. The heterogeneity of the TME is mainly due to the complex ecosystem created by the interactions among tumor cells, stromal cells and immune cells, all immersed in a dense and dysregulated ECM. Poor blood flow and crowded glycolytic tumor cells form niches characterized by reduced oxygenation, pH acidity, reduced nutrient loading, collection of anti-inflammatory cytokines and chemokine, and storage of metabolic by-products such as lactate.

Understanding the mechanisms underlying cell-cell interactions is essential for mapping the tumor microenvironment (TME), which is currently a critical aspect of improving prognosis, diagnosis, and therapies. In this context, nanotechnologies, aimed at precision medicine, pave the way for a ground-breaking approach to combating cancer. This is due to their extensive applications in detecting signature biomarkers, which are crucial steps toward early diagnosis and targeted therapeutic drug delivery.[94] Optical biosensors, particularly those based on fluorescence (FL), are becoming increasingly important in cancer research due to their enhanced detection capabilities. They represent valuable tools for detecting and analysing a wide range of biomolecules, making them advantageous for studying cancer-related processes.[95] Fluorescence microscopy is a critical tool for bio-imaging and optical sensing of specific analyte concentrations in tissues and cancer models,[96] where the most commonly used fluorescence microscopy techniques include fluorescence lifetime imaging microscopy (FLIM),[97] phosphorescence lifetime imaging microscopy (PLIM),[98] and near-infrared (NIR) microscopy.[99] While the methods mentioned earlier require specialized and sophisticated equipment, the introduction of ratiometric fluorescence (FL) measurements has significantly improved the performance of fluorescence microscopy applications.

Ratiometric FL enables the precise measurement of analyte concentrations within the TME, making it one of the increasingly valuable tools in cancer research. Many research groups focused on the realization of responsive FL molecular probes to detect single targets within the TME. For example, Anderson *et al.* developed a ratiometric pH-sensitive fluorescent dye based on seminaphtharhodafluor (SNARF) core to compare the surface cell pH of cancer cells grown either in spheroids or in mouse tumor models or in excised tumors.[100] In a different approach, Zheng *et al.* employed an iridium-based hypoxia-activated optical molecular probe to produce an oxygen nanosensor suitable to perform hypoxia imaging in mice bearing hepatoma cells H22.[101,102] Recently, the possibility of detecting two cancer parameters at the same time within the TME has provided the opportunity to deeper correlate the cancer hallmarks each other. In the same research group, Zheng and collaborators designed and synthesized an ultrasensitive molecular probe based on a poly(ethylene glycol)-conjugated iridium (iii) complex (Ir-Im-PEG). The imaging of tumor acidity and hypoxia, carried out simultaneously, was studied and performed both *in vitro*, using HeLa cell line, and *in vivo* by implanting subcutaneously H22 tumors in mice.[103] In a different work, Yeh and co-workers successfully employed commercially available probes SNARF-1 and Rhod-5N to map and to quantify protons and calcium concentrations, by means of an imaging approach. The data obtained allowed the correlation of pH and calcium levels in the intravascular and in the interstitial space of bone marrow in mouse calvarium.[104] Thus, sensing FL analytical platforms are among the most used tools since they permit the investigation of the physiological and pathological processes of living organisms.[105–107] Many of them have been developed and act as indicators for the monitoring and the quantification of specific analytes and clinically relevant metabolites.[108,109]

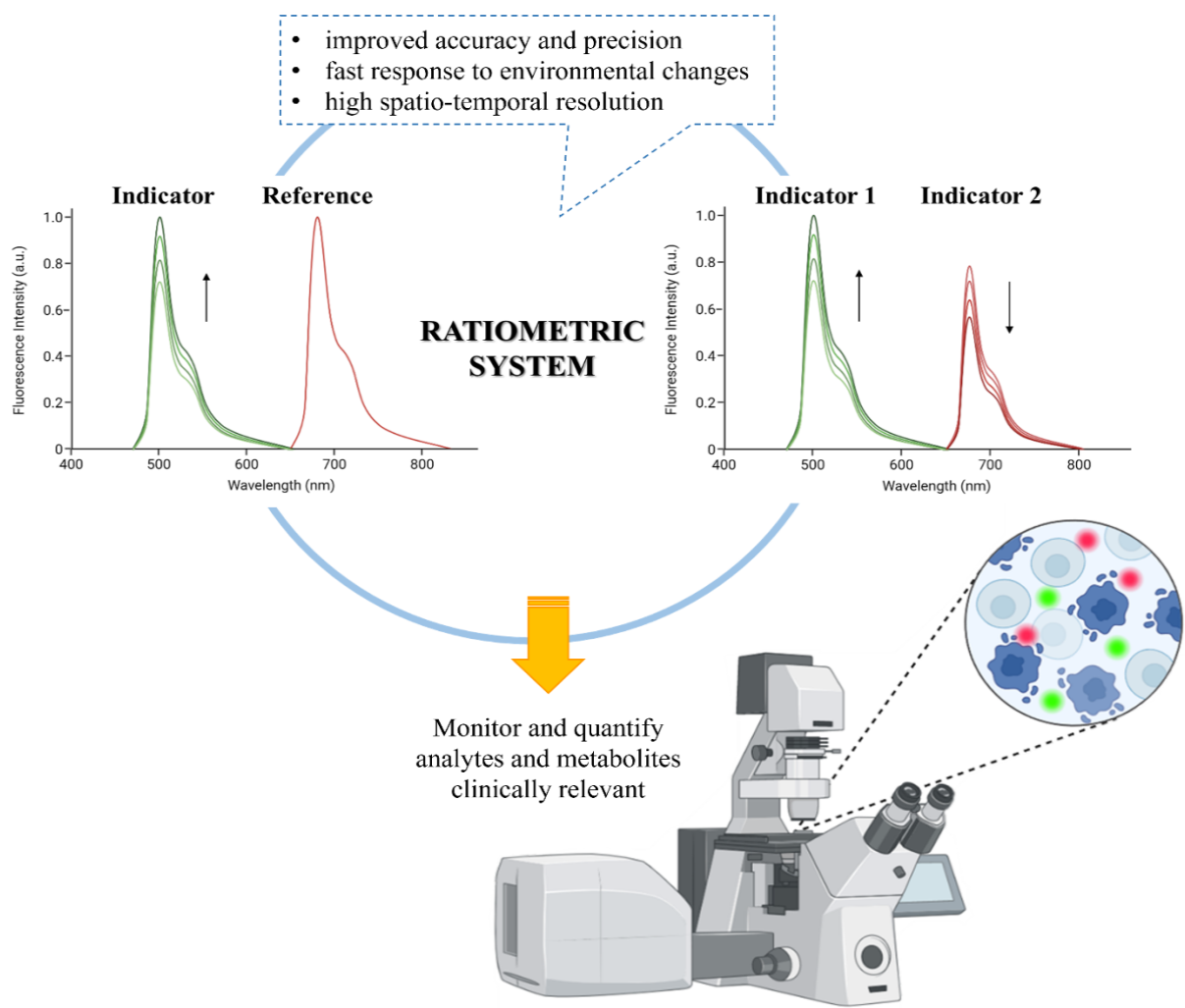

**Fig. 2** Sketch of the ratiometric optical methods with their advantages and applications in sensing TME *in vitro* and *in vivo*.

Despite advancements in fluorescence based sensor technology, designing sensors for *in vitro* and *in vivo* applications remains a challenging task due to numerous factors to consider, such as sensor selectivity, sensitivity, biocompatibility, and stability.[110] The vast potential of particle-based systems in elucidating the intricacies of the TME is due to the infinite possibilities of tuning materials, size, shape, surface charge, and functionalities, coupled with their ease of preparation and, in some cases, intrinsic biocompatibility. Silica ($SiO_2$), polystyrene (PS), and poly(methyl)methacrylate (PMMA) nanoparticles are particularly noteworthy due to their large specific surface area, stability properties, and viability. These features make them attractive candidates for *in vitro* and *in vivo* applications in cancer research.[111,112] Additionally, the coupling of fluorescence (FL) molecular probes to structured micro- and nano-systems has garnered significant interest due to their high analyte sensitivity, low cost, and rapid spatiotemporal resolved measurements. The unique features of nanoparticles make them ideal for studying various cancer-related processes, such as bio-imaging and optical sensing of oxygen concentration in tissues and cancer models using fluorescence microscopy, including FLIM. Organic dyes with specific excitation and emission wavelengths can serve as probes to reliably respond to specific targets in the TME, making them valuable tools for cancer research. The FL behaviour of the molecular probe is determined by structural changes of the fluorophores, which result from the formation and/or breakdown of responsive functional groups. These changes can cause FL quenching (off) or FL enhancement (on) effects,[113] which are the results of Förster Resonance Energy Transfer (FRET), photoinduced energy transfer (PET), internal charge trasfer (ICT), and self-quenching phenomena. Understanding these phenomena is crucial for developing effective molecular probes for various applications in cancer research. Although single dye-doped sensors are widely used and easily manipulated, they often lack accuracy due to instrumental, operational, and environmental variations that can interfere during analyses. To address this issue, single-signal sensors have been replaced by ratiometric FL sensors. These sensors incorporate a second FL target-insensitive signal that serves as a reference signal, reducing the possibility of errors through self-calibration and refining the sensors' detection limits. This advancement has significantly improved the accuracy and reliability of molecular probes used in cancer research.[114,115] Another strategy to build-up a ratiometric FL-sensor is the employment of a dual-emissive probe, which presents two reversible detection signals strictly interrelated (**Figure 2**). Thus, as a general rule, a ratiometric analysis is calculated by plotting the ratio of two FL signals. Moreover, the encapsulation of dyes within the matrix of biosensors can improve the performance of the developed analytical platforms in terms of target selectivity, there are still some aspects that represent current challenges, regarding environmental perturbations over long time experiments, photobleaching, and light scattering phenomena. Within the vast world of nanosized-photoluminescent particles it is possible to identify two main categories: quantum dots and metallic nanoclusters.[116] The multiple advantages observed in the field of fluorescence imaging over the last few years mainly stem from the discovery and the development of brilliant nanoparticles, synthetically obtained using elements from group II–VI, IV–VI and III–V of the periodic table via different techniques.[117–119] These nanoparticles, with quantum confinement effects, are defined as semiconductor nanocrystals, or simply quantum dots (QDs). Single properties such as size-dependent emission, narrow emission peaks, and resistance to photobleaching promote the realization of optical sensing systems based on QDs, in comparison with the use of organic dyes.[120] However, biosensing systems exploiting QDs potentials must overpass their citotoxicity in biological systems[121] by surface modifications and/or coating with biocompatible materials.[122–124] On the other side, metallic nanoclusters (MNCs), metal centred nanoparticles that can be stabilized by protective groups, usually biological molecules, represent powerful alternative platforms for fluorescence sensing *in vitro* and *in vivo* and their investigation is growing over years. Among their properties, it is important to remark intrinsic water-solubility and biocompatibility.[125,126]

In this comprehensive review, we delve into the cutting-edge developments in the detection and monitoring of critical TME parameters, including pH, $O_2$, ROS, and inorganic ions, along with crucial tumor biomarkers. Through the implementation of nano- and microparticle-based ratiometric fluorescence sensors, we present an in-depth analysis of their latest integrations in two and three-dimensional architectures. Our review includes a detailed description of the intricate mechanisms behind these sensors and showcases their impressive *in vitro* and *in vivo* applications, while providing an insightful analysis of the strengths and limitations of each system.

## 2. Microenvironment parameters under study

### 2.1 pH

The real-time monitoring of pH in biological environment is a challenging task[127] that cannot be fulfilled by standard methods. In lab practice, the pH-meter electrodes used are cheap and reliable tools for bulk pH measurements, but the difficult miniaturization of these devices makes them less suitable for *in vitro* and *in vivo* studies. From a physiological point of view, proton ($H^+$) concentration varies from one cellular compartment to another. For instance, pH in the cytosol has a value of circa 7.0-7.4 values, it is among 7.2 in the endoplasmic plasmatic reticulum (ER), while it is slightly acidic in the organelles, as it is 6.4 in the Golgi apparatus, 5.0 in lysosomes, 5.4 in secretory granules, 6.2 in early endosomes, 5.3 in late endosomes, 8.0 in mitochondria.[84,128–130] The metabolic switch induced in cancer growth breaks the balance between the cells' compartments, determining pH fluctuations, not only in the intracellular environment, but also in the extracellular surrounding. Therefore, the need of monitoring and mapping pH, inside the cells and in the space between cells, constitutes a crucial topic of research interest. With this purpose, FL small molecules and nanoprobes have been intensively developed and studied for sensing pH.[131] Optical pH measurement is based on the significant change in absorption or fluorescence of suitable pH indicators after protonation/deprotonation at different pH values. Today many fluorescent pH indicators are commercially available (e.g., fluorescein, semi-naphtharhodafluor 1 (SNARF1), 8-hydroxypyrene-1,3,6-trisulfonic acid trisodium salt (HPTS), Nile Blue A) and most of them have been successfully employed in non-invasive and real-time imaging of pH in several physio-pathological processes.[132] Nonetheless, one of the main drawback of pH-sensing molecules is their limited sensitivity range and, for some of them, lack solubility in water solutions, as well as toxicity. Therefore, the encapsulation of FL pH-sensing molecules into nano-structured and biocompatible matrices improves the final analytical platform in terms of photostability, solubility and cell viability, enhancing accuracy. A striking example are the polyelectrolyte multilayer capsules, obtained via the layer-by-layer (LbL) method, that, in the past two decades have irrupted in the scene of nanotechnology as a straightforward and versatile technique.[133–135] In 2011, del Mercato employed the dual-emission ratiometric SNARF1-dextrane derivate to prepare permeable calcium carbonate ($CaCO_3$)-based capsules via the LbL technique.[136] The ratiometric SNARF1 dye has the unique property of displaying two emission peaks, depending on pH: the excitation at 543 nm determines, at acidic pH, an emission peak with a maximum value at 594 nm, while, in basic pH, a spectral emissive band at 640 nm is recorded. The FL characterization, employing spectroscopy and FL microscopy, carried out on the pH sensing capsules, confirmed the ability of the labelled and encapsulated amino-dextran SNARF ($\lambda_{ex}$= 543 nm; $\lambda_{em1}$= 594nm; $\lambda_{em2}$=640 nm) to efficiently sense $H^+$ concentrations with the same sensitivity of the free dye. Later, the same capsules were applied for measuring intracellular pH in MCF-7 breast cancer cells. The cellular uptake of capsules was tracked to monitor pH in the endosomes and lysosomes.[137,138] In a different approach, multilayer pH-sensing capsules, based on SNARF probe, were successfully applied to map the pH microenvironment of human mesenchymal stromal cells (hMSCs) seeded in 3D additive manufactured scaffolds.[139] A fully automated computational approach for precisely measuring organelle acidification in cancer cells was set up by Chandra and colleagues.[140] The authors developed micrometer-sized silica ($SiO_2$) particles that were functionalized with fluorescein 5-isothiocyanate (FITC) ($\lambda_{ex}$= 492 nm; $\lambda_{em}$= 518 nm), as pH probe, and rhodamine B isothiocyanate (RBITC) ($\lambda_{ex}$= 570 nm; $\lambda_{em}$= 595 nm), as the reference dye. Furthermore, to ensure the internalization in the cytosolic comportment, the MPs were decorated with a net positive external charge which allowed cell uptake as well as the finest acidic sensing property because of the lowered pKa (6.30 ± 0.09). The tracking and mapping experiments were performed by means of CLSM time-lapse, using as tumor models MDA-MB-231 and MCF-7 breast adenocarcinoma cell lines (**Figure 2.1 a**). The innovation of the entire method resided in the automated computational approach that simplified and enriched the interpretation of data derived from image acquisitions through the creation of *ad hoc* algorithms.

Over the last decade, many groups have developed new FL probes having a wider range of sensitivity towards pH values. This is the case of Srivastava and collaborators,[141] who recently synthesized a novel pH-responsive green naphtalimide-based dye ($\lambda_{ex}$= 405 nm; $\lambda_{em}$= 525 nm) (**Figure 2.1 b**). The dye was obtained by covalently linking two functional moieties: a selective lysosomal targeting part, represented by a morpholine unit bound to 4- bromo-1,8-naphthalic anhydride, and a piperazine ring, which improves the solubility of the dye in water. The working mechanism of the pH-sensing dye was studied over all pH ranges. In detail, the protonation of the morpholine and piperazine amine groups determined the switch-on in an acidic environment of the green FL signal, which therefore was PET-induced quenched gradually passing across neutral and basic pH, thus characterizing the probe selectivity range between 2.0 and 8.0 pH values. The plus of this work consisted in the fabrication of ratiometric $SiO_2$ NPs, coupling the reference RBITC dye ($\lambda_{ex}$= 570 nm; $\lambda_{em}$= 595 nm) to the green synthesized pH-indicator dye, and their following *in vitro* application for mapping the lysosomal uptake and pH fluctuations, by means of CLSM, in human lung cancer A549 cell line. While the developed analytical platform has demonstrated its suitability for this purpose, the authors acknowledge the need for deeper live-cell imaging studies, specifically regarding the co-localization of the NPs in the endosomes and lysosomes. Such studies are crucial to further validate the effectiveness of this innovative approach in cancer research.

Although pH-sensitive fluorescent nanosensors based on hydrophobic indicators are largely unexplored, boron-dipyrromethene (BODIPY) and boron-azadipyrromethene (aza-BODIPY)-based dyes are noteworthy examples. The introduction of hydrophilic moieties to the BODIPY core has significantly enhanced their water solubility while retaining their fluorescence properties.[142] Currently, near infrared (NIR) emissive aza-BODIPY pH-indicator compounds were synthesized by Strobl *et al.* and were presented as novel dyes covering the pH scale from 1.5 to 13.[143] The great advantages in using such type of dyes in long-wavelengths spectral region are the enhanced photostability, less scattering background and deep light penetration. Despite several studies *in vitro* and *in vivo* of BODIPY dyes,[144] still biotechnological development of ratiometric pH platforms and their further applications in cell sensing are not present in literature. However, the successful engagement of hydrophobic pH sensing dye chromoionophore III (Ch3 or ETH 5053) for exploring lysosomal pH was published by Chen and collaborators in 2022 (**Figure 2.1 c**).[145] In this work, the authors reported the encapsulation of the hydrophobic Nile Blue A-derivate Ch3 into the polymeric matrix of poly(styrene)-graft-poly(ethylene oxide) (PS-PEO) NPs. Ch3 is a ratiometric pH sensing probe, characterized by two excitation wavelengths and emissive spectra in the far-red region ($\lambda_{ex1}$= 586 nm and $\lambda_{em1}$= 675 nm; $\lambda_{ex2}$= 469 nm and $\lambda_{em2}$= 575 nm), corresponding to the protonated and the deprotonated forms, respectively. The mechanism of FL emission is regulated by FRET phenomenon: moving from pH values from 10 to 3, the emission peak at 575 nm gradually decrease upon excitation at 469 nm, while *vice versa* the emission peak at 675 nm gradually increase. The linear regression recorded by plotting the FL intensity at 675 nm *versus* the FL intensity at 575 nm enabled the authors to test the pH nanosensors following their incubation with HeLa cells. The Ch3-NPs were endocytosed by HeLa cells and subcellular pH monitoring was carried out using time-lapse CLSM acquisitions. After proper pH calibration of sensors, carried out in cell medium, the pH of the organelles was determined to be around 4.7. The sensor system proposed by Chen and colleagues[145] greatly summarized the potential of ratiometric NPs engineering, highlighting the easy preparation through the dye embedding in a biocompatible matrix, thus making the system suitable for future *in vitro* and *in vivo* applications.

## 2.2 Oxygen

The hypoxia switch-on represents a hallmark within the TME that leads towards concomitant events such as the metabolic switch, acidosis and ECM rearrangement phenomena, which in turn are involved in progression and MDR of cancer.[146] Therefore, having a look at the variations of the partial pressure of oxygen ($ppO_2$) appears as a chief strategy to be monitored, both *in vitro* and *in vivo*.[147] Nowadays, several methods for the detection of dissolved oxygen concentration are widely published in literature and they are based mainly on electrochemical (amperometry, potentiometry, or conductometry) and chemical (Winkler titration) techniques.[148] While various methods for sensing TME parameters exist, they often lack dynamic real-time and spatiotemporal resolution at a single-cell scale. However, in recent decades, optical and ratiometric FL probes have become the preferred methodology for bioimaging applications, and the field continues to see ongoing advancements. The principle behind oxygen sensing relies on the quenching of luminescence intensity of an indicator probe by molecular oxygen ($O_2$), a phenomenon governed by energy or electron transfer mechanisms that are well described in the Stern-Volmer equation. By utilizing such probes, researchers may gain valuable insights into oxygen levels within the TME, which may have important implications for the development of novel diagnostic and therapeutic approaches in cancer research.[149] Moreover, another important parameter to take into account is the diffusion coefficient (D) of $O_2$ throughout the matrix in which the indicator is immersed. Thus, the combination of the choice of the optimal oxygen sensing indicator, together with an unquenchable and photostable reference dye, and the selection of the suitable, inert and gas-permeable/ion-impermeable matrix, makes possible the fabrication of $O_2$ ratiometric FL sensing devices. In this portrait, a vast variety of oxygen-permeable materials (silicon polymers, organic glassy polymers, fluoropolymers and cellulose derivate polymers)[150,151] and reversibly quenched oxygen indicator probes (e.g. transition metal polypyridyl complexes and metalloporphyrins)[152] are commercially available.

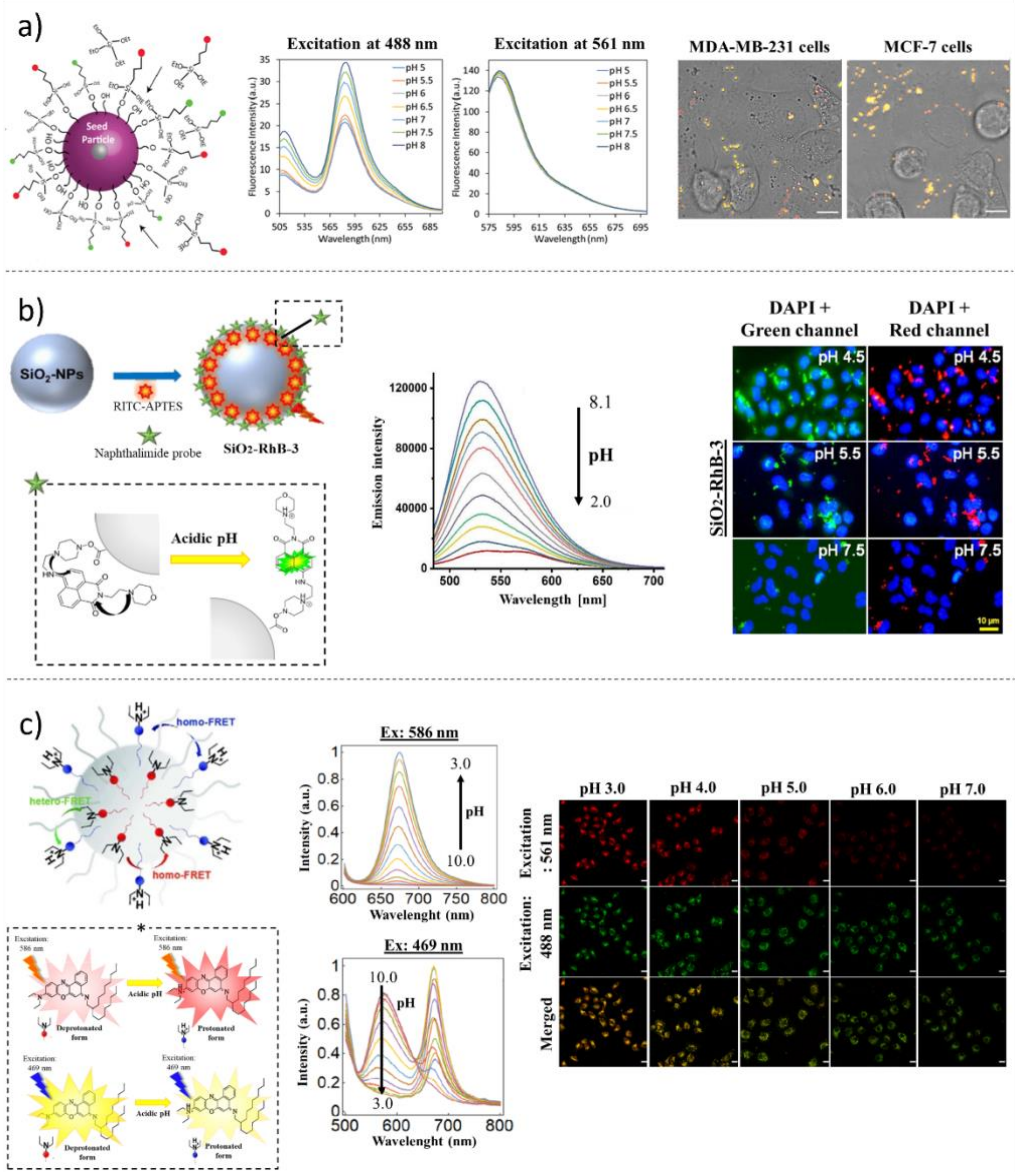

**Fig. 2. 1 Examples of pH sensing ratiometric nano-platforms**. **a)** *(left)*: schematic illustration of the ratiometric $SiO_2$ MPs functionalization with FITC and RBITC dyes by a modified Stöber method; *(middle)*: pH-dependent fluorescence of the MPs; *(right)*: fluorescence micrographs showing the color changes of the ratiometric pH-responsive MP sensors added to MDA-MB-231 cells and MCF-7 cells after 24 hours' incubation. Scale bars: 10 μm. *Adapted with permission from Chandra et al., ACS Appl. Mater. Interfaces 2022, 14, 18133–18149; figure licensed under CC-BY 4.0, https://creativecommons.org/licenses/by/4.0/*. **b)** *(left)*: schematic representation of the protocol used to synthesize pH-sensing $SiO_2$-NPs; *(middle)*: pH-dependent fluorescence of probe 3 on NPs; *(right)*: epifluorescence images of fixed A549 cells incubated with solutions of pH nanosensors $SiO_2$-RhB-3, monitoring with emission filters set to $\lambda_{em}$ = 470 nm (green channel) and to $\lambda_{em}$ = 560 nm (red channel). Scale bar: 10 μm. *Adapted with permission from Srivastava et al., Sci Rep 13, 1321, 2023; figure licensed under CC-BY 4.0, https://creativecommons.org/licenses/by/4.0/*. **c)** *(left)*: schematic illustration and working principle of the protonation of Ch3 at the surface of the NPs, turning the color from red to blue.; *(middle)*: fluorescence emission spectra (excitation at 586 nm) of the nanosensors containing Ch3, PS-PEO, and NPOE in universal buffer solutions at different pH from 10 to 3, upon excitation at 586 nm and 469 nm; *(right)*: CLSM images for the cellular pH calibrations of the nanosensors from pH 3.0 to 7.0. Scale bar: 20 μm. *Reprinted with permission from Chen et al., Nano Res., 2022, 15(4): 3471–3478. Copyright © 2021, Tsinghua University Press and Springer-Verlag GmbH Germany, part of Springer Nature.*

Among all the transition metal dyes, ruthenium (Ru(II))-based polypyridyl complexes are extensively adopted in bioimaging applications, thanks to their properties such as large Stokes' shift, excitation and emission band in the visible region, good photostability and high brightness.[153] Despite these characteristics, the excitation and the emission peaks are quite broad, which can be interpreted both positively, having more possibility of choice for the laser excitation, or negatively, because of the difficulties in isolating the emission band during multi-analyte measurements.[152] Nevertheless, Ru(II))-based polypyridyl complexes remain the most used $O_2$-molecular probes. For instance, Xu et al. reported in 2001 the first sol-gel-based ratiometric FL PEBBLEs for the real-time measurements of oxygen in rat C6 glioma cell line.[154] The nanosensor was built-up by encapsulation within the matrix of silica particles, obtained by a modified Stöber method, of the indicator and reference dyes, which were ruthenium (II)–tris(4,7-diphenyl-1,10-phenanthroline) dichloride ($[Ru(dpp)_3]^{2+}$) ($\lambda_{ex}$= 543 nm; $\lambda_{em}$= 610 nm) and Oregon Green 488-dextran ($\lambda_{ex}$= 488 nm; $\lambda_{em}$= 525 nm), respectively. The ratiometric linear regression plot, obtained from the calibration curve at a different rate of $O_2$, allowed the authors to inject the ratiometric FL nanoprobes into rat C6 glioma cells and collect preliminary information regarding intracellular oxygen thought CLSM image acquisitions. If, at that time, the novelty of the work published was represented by being the first development and application of a ratiometric $O_2$ sensing system, however the platform possessed a great limitation, that was the absolute need to control pH within the cell medium since Oregon Green is a pH-sensitive dye for pH values below 6.0.

Platinum (Pt(II))- and palladium (Pd(II))-based metalloporphyrins constitute another category of optical $O_2$ sensors having strong phosphorescence, good molar absorption coefficients and large Stokes' shifts.[155] In the work reported by Wu and collaborators, platinum(II)-octaethyl porphyrin (PtOEP) ($\lambda_{ex}$= 580 nm; $\lambda_{em}$= 650 nm) was employed as oxygen sensitive dye and entrapped by means of particle precipitation technique into the matrix of polyfluorene derivates poly(9,9-dihexylfluorene) (PDHF) ($\lambda_{ex}$= 350 nm; $\lambda_{em}$= 420 nm) and poly(9,9-dioctylfluorene) (PFO) $\lambda_{ex}$= 350 nm; $\lambda_{em}$= 420 nm), which in turn acted as reference signals in the ratiometric system and as hydrophobic, glassy and gas-permeable polymers.[156] The working mechanism of the probes was based on the FRET phenomenon, in which the PDHF and PFO were the donor units, while PtOEP was the acceptor one. In this study, π-conjugated polymer nanoparticles (CPdots) obtained exhibited peculiar properties: under excitation the polymer matrices were capable to transfer energy to the phosphorescent PtOEP, thus enhancing its sensitive ability to respond at different concentrations of oxygen; additionally, the confinement inside the polymeric matrices determined an augmented photostability. So, the fluorescence emission of the PtOEP was linearly quenched in presence of increasing concentrations of $O_2$, while PDHF and PFO did not change their fluorescence signal. Following the characterization of the sensors, the authors tracked the cellular uptake operated by macrophage-like murine J774A1 cells line using the differential interference contrast (DIC) images and phosphorescence images of the nanoparticle-labelled cells, demonstrating the possibility to detect subcellular $O_2$ concentrations. The overall results indicated that CPdots possessed great potential in the quantification of oxygen in in vitro experiments dictated by their unique qualities such as brightness, ratiometric emission, small size and, in consequence, cellular uptake. Taking advantage of these findings, in 2012 in Wang's lab,[157] PtOEP was engaged to prepare a ratiometric FL probe, together with the $O_2$-insensitive dye coumarin 6 (C6) ($\lambda_{ex}$= 381 nm; $\lambda_{em}$= 510 nm), fabricated by precipitation and encapsulation of the dyes within the matrix of PMMA nanoparticles, whose surface was functionalized with poly-L-lysine to facilitate cellular uptake and thus intracellular oxygen imaging visualization (**Figure 2.2 a**). The linear regression obtained by the Stern-Volmer plot, and the characterizations performed, allowed the authors to apply the nanoprobes in cellular cultures of HepG2 human hepatocellular liver carcinoma cell line. CLSM images were acquired after 12 hours of incubation of the NPs with cells in culture medium. Assuming a vesicular cellular up-take, the NPs were found to be localized in the intracellular space of HepG2 cells. Pushing the cell cultures towards hypoxia and normoxia conditions, Wang and collaborators could check the goodness of the developed ratiometric FL $O_2$ analytical platform, confirming the potentiality of such a tool to sense intracellular oxygen concentrations.[157]

Concerning the need of improving the existing $O_2$-sensing probes, later Xu et al. drew up a ratiometric metal-organic framework (MOF) for sensing intracellular oxygen (**Figure 2.2 b**).[158] MOFs are nanomaterials synthesized by bridging metal ions or clusters with the help of organic ligands, which can be molecular sensing probes, thus preparing a final unit with a specific analytical scope.[159] For this reason, in the last years, researchers explored MOFs as promising devices because of their intrinsic characteristics: firstly, encapsulated dyes and therapeutic drugs can diffuse easily from the core thanks to the high porosity of the material; secondly, the encapsulation of sensing molecules permits to enhance the photostability and to reduce the self-quenching phenomenon; lastly, the covalent bounds between the metal and organic linkers prevent the unforeseen leaching.[160,161] Thus, exploiting the challenging MOFs' properties, Xu

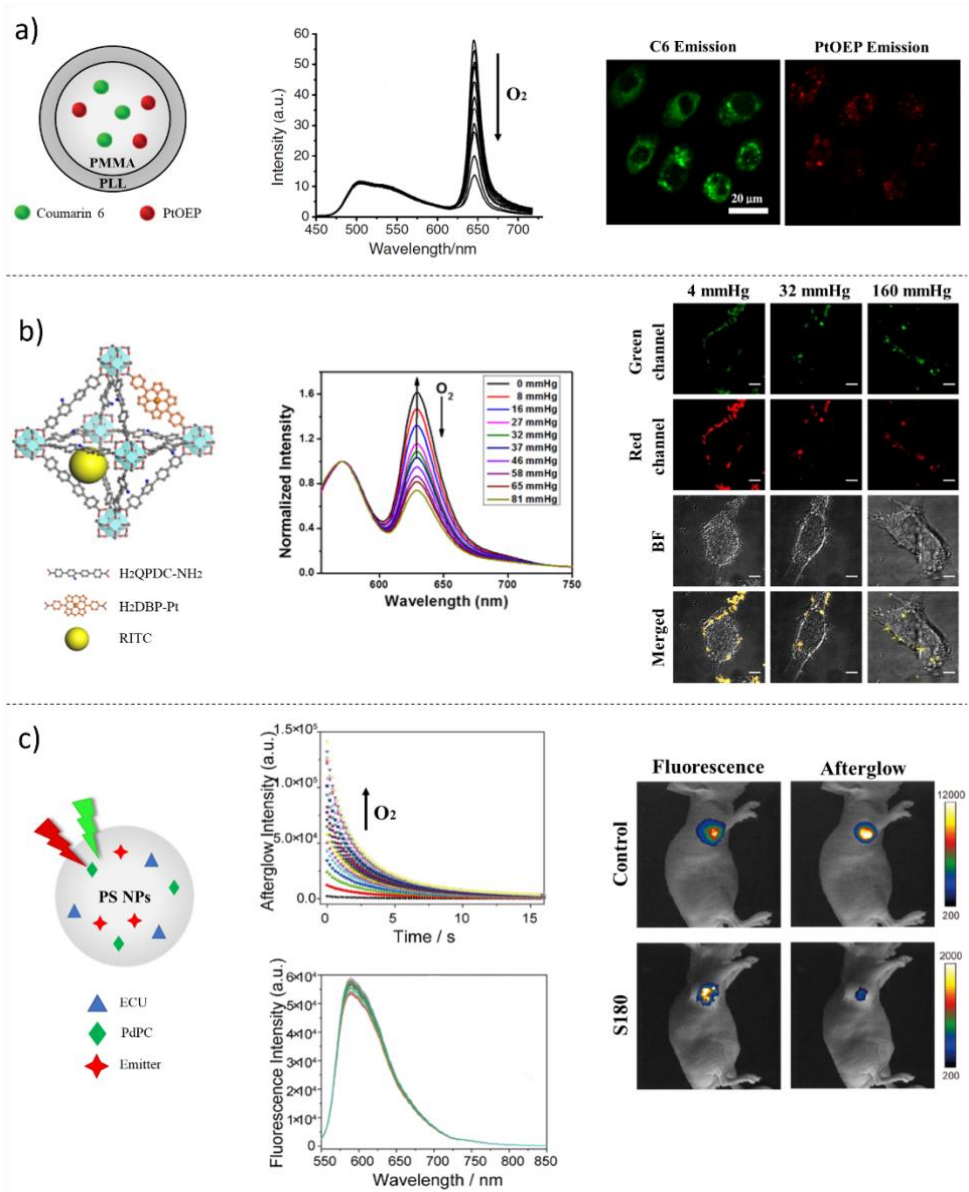

**Fig. 2. 2 Examples of O₂ sensing nano-platforms. a)** (*left*) schematic representation of the oxygen ratiometric PMMA-NPs; *(middle)*: emission spectra of the sensor NPs at various concentrations of oxygen upon excitation at 381 nm excitation; *(right)*: CLSM images of HepG2 cells loaded with the oxygen sensing NPs in normoxia conditions; the green fluorescence of C6 of ratiometric NPs was recorded using a 560 nm emission band-pass filter with a 405 nm excitation, while the red fluorescence of PtOEP using a 750 nm emission band-pass filter with a 543 nm excitation wavelength. Scale Bar: 20 μm. *Reprinted with permission from Wang et al., Microchim Acta, 2012, 178, 147–152; Copyright ©2012, Springer-Verlag.* **b)** *(left)*: schematic representation of R-UiO NMOF based on Pt(II)-porphyrin ligand as an O₂-sensitive probe and a Rhodamine-B isothiocyanate ligand as an O₂-insensitive reference probe; *(middle)* emission spectra ($\lambda_{ex}$ = 514 nm) of R-UiO in HBSS buffer under various oxygen partial pressures; *(right)*: CLSM images of CT26 cells under hypoxia (4 mmHg), normoxia (32 mmHg), and aerated conditions (160 mmHg) after incubation with R-UiO-2. Scale bar: 5 μm. *Reprinted with permission from Xu et al., J. Am. Chem. Soc. 2016, 138, 2158–2161. Copyright © 2016, American Chemical Society.* **c)** *(left)*: schematic illustration of the working principle of afterglow/fluorescence dual-emissive ratiometric O₂ probe; *(middle)*: afterglow decay curves of AGNPs and fluorescence spectra of AGNPs in different oxygen concentrations; (right): fluorescence and afterglow images of mice with the subcutaneous implantation of AGNPs in the mouse bearing no tumor and of mice with the intratumor injection of AGNPs in the mouse bearing the tumor, and the corresponding fluorescence and afterglow intensity. *Reprinted with permission from Wen et al., Anal. Chem. 2023, 95, 4, 2478–2486. Copyright © 2023, American Chemical Society.*

R. and co-workers prepared a nanomaterial R-UiO-based having Pt-5,15-di(*p*-benzoato)porphyrin (H$_2$DBP-Pt) ($\lambda_{ex}$= 570 nm; $\lambda_{em}$= 595 nm) as bridging ligand, sensible to changes in O$_2$ concentration, and RBITC-conjugated quaterphenyldicarboxylate ($\lambda_{ex}$= 570 nm; $\lambda_{em}$= 595 nm), as reference fluorophore. The authors recorded a linear Stern-Volmer regression when the R-UiO was calibrated in presence of different concentrations of O$_2$., Mouse colon carcinoma CT26 cell line was chosen as a cancer model, and Hank's balanced salt solution (HBSS) was used as buffer medium for measuring O$_2$ intracellular levels. Through CLSM imaging, an efficient cellular uptake was tracked and captured with image acquisitions after 2 hours of incubation of R-UiO MOFs with CT26 cells. The authors carried out the experiments using three O$_2$ concentrations: hypoxic (4 mmHg), normoxic (32 mmHg) and aerated (160 mmHg) conditions. The ratiometric calculations, obtained via image analysis of the internalized R-UiO, corresponded to 5.1 ± 2.5, 27.3 ± 3.1, and 158 ± 11 mmHg, data that matched with the theoretical O$_2$ values, demonstrating the accuracy of the proposed method. A step forward in imaging oxygen at tissue levels has been done in a reported work of early 2023 published by Wen *et al.*, in which an afterglow/fluorescence dual-emissive ratiometric O$_2$ sensor was engineered (**Figure 2.2 c**).[162] In particular the probe was based on a photochemical reaction-based afterglow system: the afterglow palladium (II) 1,4,8,11,15,18,22,25-octabutoxyphthalocyanine (PdPc) ($\lambda_{ex}$= 730 nm; $\lambda_{em}$= 600 nm) was the indicator dye sensitive to variations of O$_2$ concentration, while Nile Red (NR) ($\lambda_{ex}$= 582 nm; $\lambda_{em}$= 600 nm) represented the unquenchable O$_2$ dye. The reaction at the basis of the ratiometric FL system happened as followed: in presence of O$_2$, the FL intensity of the afterglow gradually increased because of the formation of singlet oxygen species, while the reference NR kept its FL peak unperturbed. The encapsulation of the fluorophores within polystyrene particles (AGNPs) through the swelling method allowed dyes confinement, oxygen penetration, good solubility and viability.[162] The plus described in the research article of Wen and collaborators consisted in being among the firsts in coupling afterglow/fluorescence sensing procedure for *in vitro* and *in vivo* experiments to explore hypoxia environment in solid tumors. To this purpose, mouse sarcoma S180 cell line was employed to test cytotoxicity of the nanoparticles performing 3-(4,5-dimethylthiazol-2-yl)-2,5-diphenyltetrazolium bromide (MTT) assay. Finally, female athymic nude mice bearing mouse sarcoma cell-derived S180 and female athymic nude mice no tumor bearing were treated with AGNPs. After two weeks, the mice were imaged showing a remarkable afterglow enhancement in mice with no tumor masses, while the afterglow signal was quite off in tumor bearing mice. Thus, in this way the quantification of the O$_2$ concentration, by means of the ratio of the afterglow signal versus fluorescence intensity, in solid tumor in the area of injection of the AGNPs showed a 4.94-fold lower intensity in comparison to the ratio obtained in normal tissues. The study of Wen et al. proposed to the scientific community a stable and accurate ratiometric sensor for O$_2$ quantification *in vitro* and *in vivo*, but some efforts are still needed to improve the quantification of oxygen in deeper solid tumors.

**2.3 Reactive Oxygen Species (ROS)**

As mentioned previously, the most important ROS are hydrogen peroxide (H$_2$O$_2$), its reduction product hydroxyl radical (•OH), and superoxide anion radical (O$_2^{·-}$).[70–72] In the complex landscape of TME, nanotechnologies play a fundamental role for ROS detection and monitoring. As consequence, in the last two decades, many efforts have been spent to develop fluorescent probes for ROS sensing.[163,164] Two are the mechanisms by which non-fluorescent ROS-sensitive dyes are activated: by H$_2$O$_2$ selective cleavage or by oxidation. In 2008, Srikun and co-workers[165] published an internal charge trasfer (ICT)-based approach to detect H$_2$O$_2$ in living cells using Peroxy Lucifer-1 (PL1) as a ratiometric fluorescent reporter. PL1 presents a 1,8-naphtalimide core structure with a 4-boronate-based carbamate protecting group that, once excited in absence of H$_2$O$_2$, displays a blue emission peak ($\lambda_{ex}$=375 nm; $\lambda_{em}$=475 nm). When in presence of H$_2$O$_2$, PL1 loses the boronate-based carbamate protecting group by chemoselective cleavage returning the green fluorescent aminonaphthalimide ($\lambda_{ex}$=435 nm; $\lambda_{em}$=540 nm). The biocompatible and ratiometric reactive dye PL1 was used to detect the endogenous concentration of H$_2$O$_2$ in RAW264.7 macrophages and in HEK 293 T cells. Some years later, Kim's group[166] prepared silica nanoparticles decorated with PL1. The aim was to develop a new scaffold as a promising tool for the monitoring of hydrogen peroxide. Nowadays, nor *in vitro* or *in vivo* applications related to PL1-SiO$_2$ particles are reported in the literature. The most used probe for the detection of ROS and oxidative stress in the cellular systems is 2',7'-dichlorofluorescein (DCF) dye. The enzymatic activity of esterases in the cytosol can convert the non-fluorescent diacetate form of 2',7'-dichlorofluorescein (DCF-DA) in its hydrophilic form generating a strong fluorescence response ($\lambda_{ex}$=488 nm; $\lambda_{em}$=525 nm) deriving, in a particular manner, from the oxidation procured by H$_2$O$_2$ among other ROS. Kim and collaborators[167] developed a ratiometric nanoPEBBLE sensor to quantitatively estimate the H$_2$O$_2$ generation from stimulated RAW264.7 macrophages *in vitro*. After DCF-DA encapsulation into Ormosil nanoparticles matrix, the nanoprobe surface was functionalized with a reference dye, Alexa568 N-succinimidyl ester, and with a membrane penetrating peptide, cysteine terminated TAT peptide, which guided the delivery of the sensing PEBBLEs directly into the cytosol. In the lab of Kazakova,[168] the amphiphilic dye dihydrorhodamine 123 (DHR123) was employed as a sensing H$_2$O$_2$ unit in the fabrication of novel lactate microcapsule sensors. DHR123 is a non-emitting molecule, but in presence of H$_2$O$_2$ it undergoes oxidation generating the green emitting rhodamine 123 ($\lambda_{ex}$=488 nm; $\lambda_{em}$=550 nm). The novelty reported by Kazakova consisted in the concept of enzyme-assisted substrate sensing. In fact, the encapsulation of lactate oxidase (LOx, an enzyme that catalyses the transformation of L-lactate into pyruvate and H$_2$O$_2$)[169,170] coupled with the embedding of the amphiphilic ROS-sensitive fluorescent dye DHR123 onto the surface of calcium carbonate (CaCO$_3$) capsules, via LbL deposition of oppositely charged polyelectrolytes, enabled to monitor H$_2$O$_2$ over time using an optical approach. The data reported in this work confirmed that the increased fluorescence of DHR123 fluorophore ($\lambda_{ex}$=488 nm; $\lambda_{em}$=550 nm) is linearly correlated to the enzymatic activity of LOx and lactate concentration in the millimolar range. Results here obtained depicted the possibility of indirectly measuring lactate in the physio-pathological ranges and, by the further implementation of the following technique, paved the future monitoring of metabolites *in vitro* or *in vivo*.

Hydroxyl radicals (•OH) are recognized as the most dangerous free radicals among ROS, since it comes from $H_2O_2$ reduction in metal-catalysed Fenton chemistry, involving free iron ($Fe^{2+}$) ions. Liu et al.[171] developed a dye doped-ratiometric fluorescent probe coupling the fluorescence response of coumarin-3-carboxylic acid (CCA), as indicator dye for •OH detection and quantification, and 6-carboxy-X-rhodamine N-succinimidyl ester (ROX-SE), as a reference fluorophore (**Figure 2.3 a**). The blue enhanced fluorescence emission of CCA is strictly linked to its reaction with •OH, which transforms CCA into 7-hydroxy coumarin 3-carboxylic acid product. The engineered ratiometric silica nanoparticles displayed a dual-emission fluorescent spectrum employing a single excitation wavelength ($\lambda_{ex}$=395 nm; $\lambda_{em}$=555 nm-620 nm). After the analytical method was validated, the authors monitored the nanoparticle uptake and detected •OH levels in HeLa cells by live imaging. Importantly, cells incubated over time with 100 µM of •OH, in presence of the sensing probe, displayed an increased blue fluorescent signal of CCA, whereas no changes in the red fluorescence of ROX could be observed. Fluorescent sensing dyes, used as reporter and imaging agents, can be classified as FRET-based molecules when changes in the electronic interactions between a donor and an acceptor happen.[117] In analytical chemistry and in live imaging applications, this concept is taking more space in the scientific literature since permits a selective detection of target analytes of particular biological importance, such as in the case of ROS. Diselenide bond (Se-Se) represents a highly reactive linker with specific selectivity towards ROS, since it is oxidized into the selenic acid. Deepagan et al.[172] successfully prepared an $H_2O_2$ on-off nanoprobe for in vitro live imaging application (**Figure 2.3 b**). This nanosensor was based on the ability of gold nanoparticles to inhibit, in a distance-dependent manner, the fluorescence of surface-bound fluorescein isothiocyanate (FITC) dye via FRET, in absence of $H_2O_2$ ("off" state). When $H_2O_2$ concentration increased, the diselenide bond broke down selectively releasing the free dye that could enhance its fluorescence ("on" state). The fluorescence characterization, performed in presence of $H_2O_2$, demonstrated that the probe enhancement was linearly dependent to the concentration of hydrogen peroxide and allowed the authors to test the sensors in activated RAW264.7 macrophages to explore intracellular $H_2O_2$ changes over time. Other reactive groups, i.e. thiochetals,[173] phenylboronic acids and thioethers,[174] have been engaged as FRET-based sensors, since they represent a nucleophilic anchor and a cleavable site for $H_2O_2$ species. An example of $H_2O_2$ sensitive FRET-based biosensor came from Feng's group.[175] In this work, the authors indirectly detected the glucose concentration coming from an enzyme-catalysed $H_2O_2$ production carried out by glucose oxidase (GOx). For the fabrication of their sensing system, a self-assembly technique was employed, using two functionalized lipophilic polymers: 4-carboxy-3-fluorophenylboronic acid (FPBA)-modified DSPE-PEG (DSPE-PEG-FPBA), and 7-hydroxycoumarin (HC)-conjugated DSPE-PEG (DSPE-PEG-HC). The addiction of Alizarin Red S (ARS) to the probe unit represented a novelty. ARS is a non-fluorescent molecule, but its ability to create adducts with boronic acids make possible the generation of a fluorescent signal. In this way, the detection approach developed by Feng et al. was based on the selective cleavage of ARS from FPBA caused by increased concentrations of $H_2O_2$. When no ARS is present in the polymeric micelles, the nanoprobe displayed a fluorescence peak in the blue region of the spectrum due to HC ($\lambda_{ex}$=405 nm; $\lambda_{em}$=450 nm). The subsequent addition of ARS to the polymeric probe determined conjugation with FPBA and consequent FRET phenomenon between ARS-FPBA adduct and HC, which generate a new peak in the red region of the spectrum ($\lambda_{ex}$=405 nm; $\lambda_{em}$=600 nm). A ratiometric fluorescence response could then be recorded in presence of $H_2O_2$. In this condition, ARS was decoupled from the probe, gaining a decrease of the emission peak at $\lambda_{em}$=600 nm and an increase of fluorescence for the peak at $\lambda_{em}$=450 nm. As a future perspective, the intrinsic potential of the multifunctional polymeric fluorescent probe proposed by Feng and collaborators can contribute to the biochemical studies of cell microenvironment.

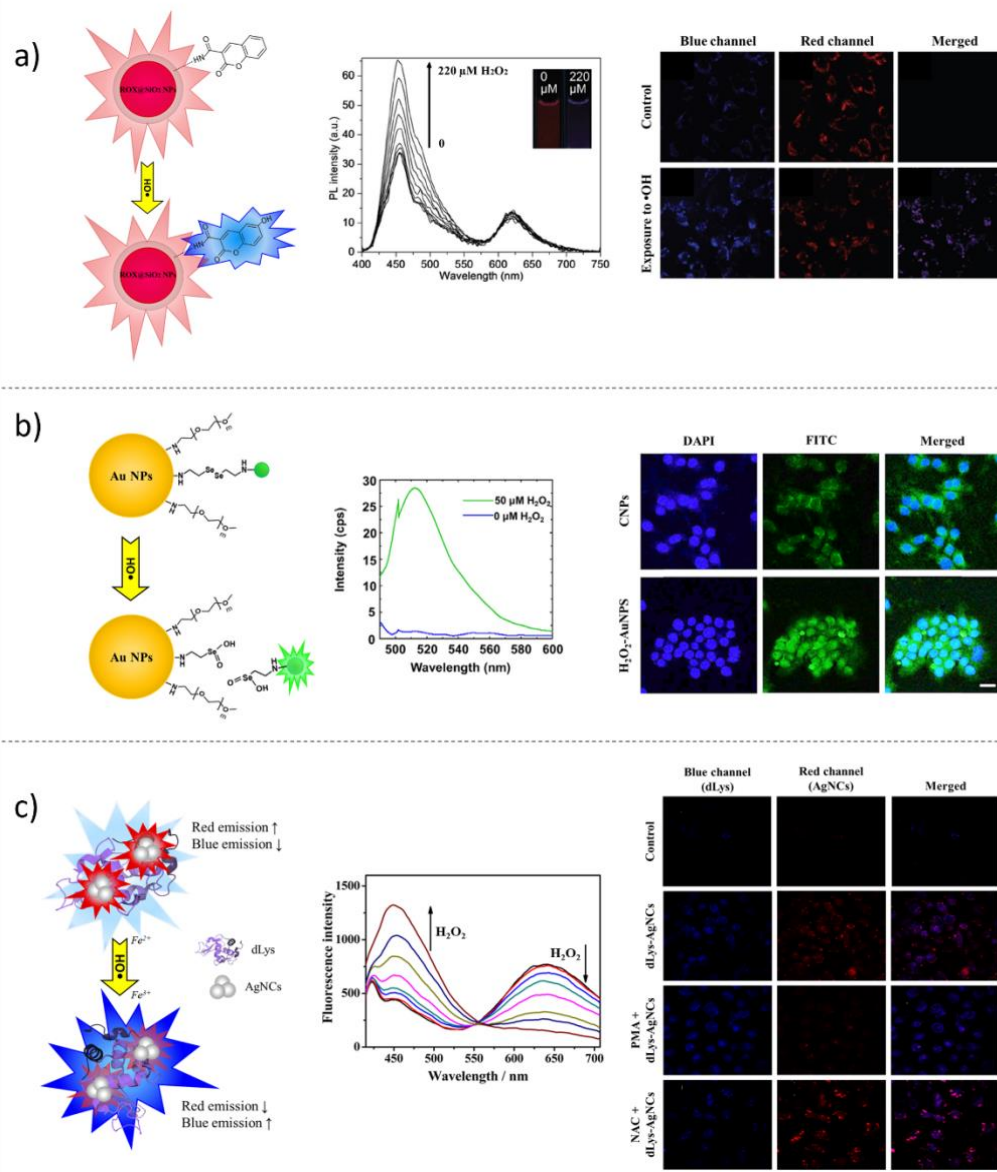

**Fig. 2. 3 Examples of ROS sensing nano-platforms. a)** *(left):* schematic illustration of the dual-emission probe synthesis procedure and the working principle for ratiometric fluorescence detection of •OH; *(middle):* fluorescence spectra of the ratiometric probe solution upon the exposure to different concentrations of •OH at various $H_2O_2$ concentrations; *(right):* confocal fluorescence images of HeLa cells after being incubated with the dual-emission probe in the absence and in presence of •OH. The images were collected at 410-520 nm (blue channel) and 580-680 nm (red channel) upon excitation at 405 nm. Scale bar: 20 µm. *Reprinted with permission of Royal Society of Chemistry, from Liu et al., Analyst 2016, 141, 7, 2296-2302; permission conveyed through Copyright Clearance Center, Inc.* **b)** *(left):* schematic illustration of the $H_2O_2$-sensitive on-off $H_2O_2$-AuNPs; *(middle):* $H_2O_2$ responsive fluorescence spectra of $H_2O_2$ sensitive–AuNPs; *(right):* in vitro confocal microscopic images of activated RAW264.7 cells incubated with the CNPs and $H_2O_2$–AuNPs for 3 hours at pH 7.4. Scale bar: 20 µm. *Reprinted from Deepagan et al., Macromol. Res., 2018, 26(7), 577-580. Copyright © 2018, The Polymer Society of Korea and Springer Science Business Media B.V., part of Springer Nature* **c)** *(left):* schematic illustration of the fluorescent responding mechanism of dLys-AgNCs to Fenton Reagents; *(middle)*: fluorescence spectra of the ratiometric NPs towards different $H_2O_2$ concentrations; *(right):* fluorescence confocal images of PC-3 cells alone (first raw), PC-3 cells treated with dLys-AgNCs probe (second raw), PC-3 cells incubated with PMA (third raw) and NAC (forth raw) prior to treatment with dLys-AgNCs. *Adopted from ref. Liu et al., Anal. Chem. 2016, 88, 21, 10631–10638. Copyright © 2016 American Chemical Society.*

QDs have been also successfully employed for $H_2O_2$ dectection. For intance, Zhou et al.[176] described the preparation and application of TGA-capped Si-CdTe dual-emissive QDs for the selective monitoring of $H_2O_2$ in the intracellular space of HeLa cells. The coupling of Si-QSDs with CdTe QDs represented an innovation not only because of the addiction of a nanosized fluorescent unit, but also because it skips the cytotoxicity problem using a silicon-based material, which has excellent solubility in water, great stability, and it is easy to obtain. The FRET-sensing system reported by Zhou's group was based on the fluorescence intensity of blue Si-QDs ($\lambda_{ex}$=370 nm; $\lambda_{em}$=442 nm) and red CdTe QDs ($\lambda_{ex}$=370 nm; $\lambda_{em}$=562 nm). By gradually increase of $H_2O_2$ concentration, the TGA-capping broke down and consequently CdTe QDs fluorescence was reduced while the Si-QDs emission became stronger, in a linear relationship. By CLSM analyses of HeLa cells cultured with Si-CdTe QDs, the authors could establish a valid method to monitor $H_2O_2$ levels following nanoparticles uptake. Metallic nanoclusters (MNCs) are metal centred nanoparticles stabilized by protective groups, usually biological molecules, and for this reason their employment as novel platforms for fluorescence sensing in vitro and in vivo is growing over years. Among their properties, it is important to remark water-solubility and biocompatibility.[125,126] Liu et al. fabricated a dual-emissive fluorescent ratiometric probe for $H_2O_2$ and •OH sensing based on lysozyme-capped silver nanoclusters (dLys-AgNCs) (**Figure 2.3 c**).[177] The fluorescent ratio, obtained between the quenching of the red peak at 640 nm and the enhancement at 450 nm due to •OH-induced oxidation of the tyrosine residue present in lysozyme, was linearly correlated to the $H_2O_2$ concentration. Live imaging experiments, performed using PC-3 cells after incubation with dLys-AgNCs, confirmed the ratiometric measurements. As a proof of concept, to the sensing system composed of PC-3 cells and dLys-AgNCs probe, the authors added phorbol-12-myristate-13-acetate (PMA) and N-acetyl-cysteine (NAC), an •OH generation stimulator and a free radical scavenger, respectively. The fluorescence confocal analysis strengthened the ability of dLys-AgNCs to detect and monitor changing levels of •OH in vitro.

## 2.4 Inorganic cations and anions ($Ca^{2+}$, $Na^+$, $K^+$, $Cl^-$)

The leading phenomena that characterize the TME, as acidosis, hypoxia and ROS generation, drag with them also the dysregulation of ions fluxes, contributing to the enhancement of the cancer disorder and strong perturbations inside cells and in their surround. Involved in cancer processes as proliferation, invasion and metastasis, ions $Ca^{2+}$, $Na^+$, $K^+$ and $Cl^-$ are the main players of cell-cell interactions and ECM digestions favouring the activation of specific signalling pathways.[178]

Calcium cations, $Ca^{2+}$, are involved in many cellular processes and signalling pathways, particularly it is essential in muscle contraction, osteogenesis and neurotransmission.[179] Its concentration is kept constant in body fluids, which it is around 100 nM in the cytosol and way higher in blood and interstitial fluids, around 2 mM.[180] These concentration differences should be kept in consideration for designing intracellular or extracellular sensors. There are different FL probes for calcium cations imaging, and some of them are commercially available such as Fura dyes, Indo dyes and Fluo dyes. All these compounds contain multiple carboxylic functions that act as chelator for $Ca^{2+}$, in an "EDTA-like" motif, with different selectivity for other bivalent cations.[80,181] Si et al. reported the development of PEBBLE (Photonic Explorers for Bioanalysis with Biologically Localized Embedding) ratiometric nanosensor for $Ca^{2+}$ imaging. The nanosensor was obtained by embedding the $Ca^{2+}$ sensitive dye, Rhod-2, in polyacrylamide nanogels. To obtain a ratiometric probe, the $Ca^{2+}$ insensitive dye Hilyte™ 647 was covalently conjugated to the nanogel surface. These PEBBLEs showed a $K_d$ of 500-600 nM and were tested on PC-3 human prostate cancer cells for intracellular live imaging.[182] Lin et al. reported the synthesis and characterization of fluorescent $Ca^{2+}$ nanosensors based on CDs. Fluorescent CDs were synthesized by pyrolysis of citric acid, purified by means of dialysis and covered with a $Ca^{2+}$ binding peptide (**Figure 2.4 a**). The peptide sequence was based on the EF-Hand domain of the human calmodulin. The binding of $Ca^{2+}$ cations to the nanosensors was able to quench the CDs fluorescence, showing a sensitivity range in the micromolar range of concentration.[183] The toxicity of the nanosensors was evaluated on SH-SYS5 human neuroblastoma cells through MTT assay, proving the low toxicity of the nanoprobes. Shulz et al. reported the synthesis of $Ca^{2+}$ cations nanosensors based on hybrid silica-dextran nanoparticles.[183] Silica nanoparticles were obtained by a modified Stöber method, doped with aminopropyl triethoxysilane (APTES) and covalently linked to RBITC. Fluo-4 modified aminodextran was used to form the outer shell around silica. The conjugation was performed by using disuccinimidyl carbonate as a linker between aminodextran and amino-modified nanoparticles. The ratiometric sensor showed a sensitivity range from 0 to 39.7 μM of $Ca^{2+}$, with a $k_d$ of 780nM, which can be suitable for intracellular $Ca^{2+}$ imaging.

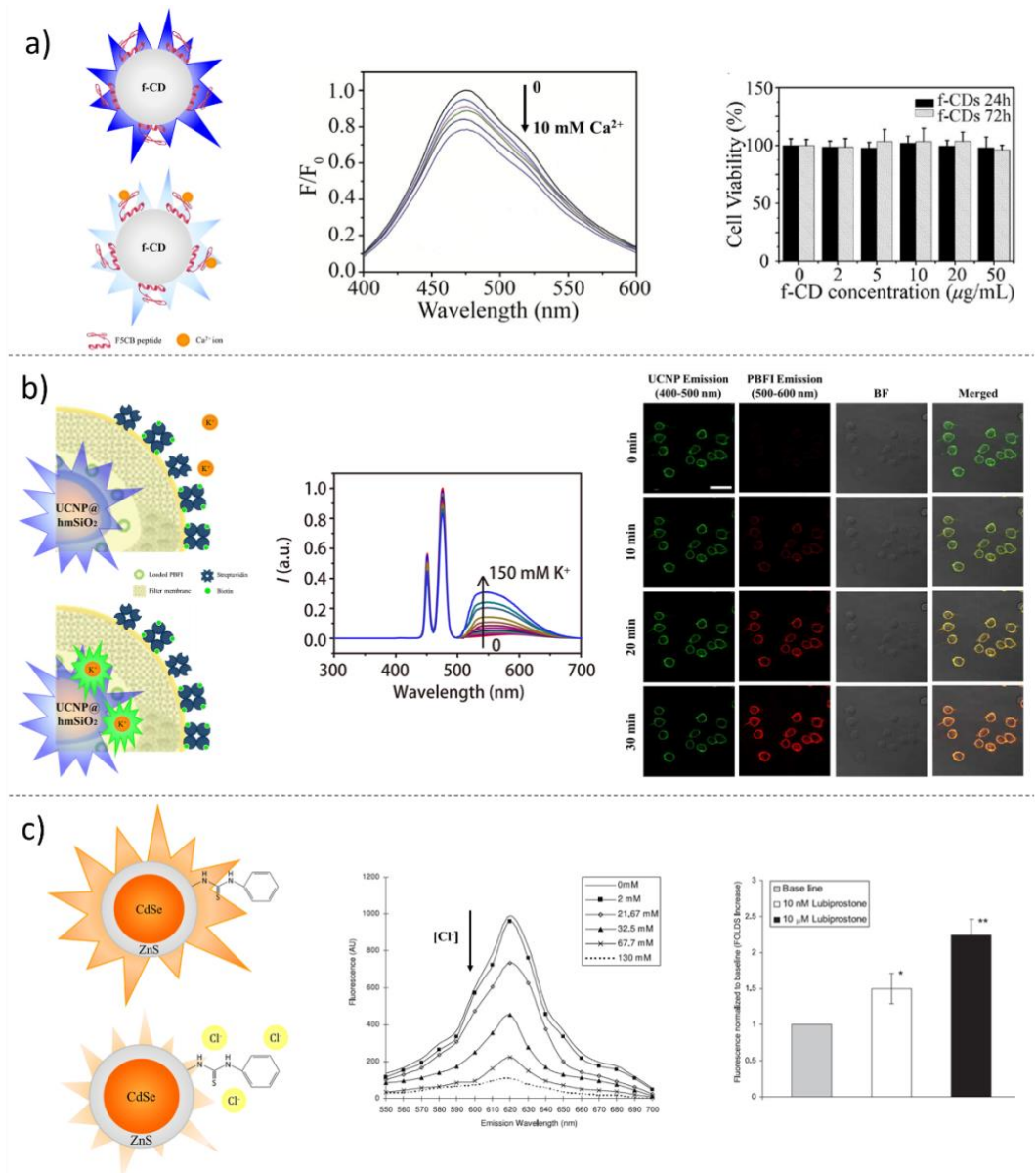

**Fig. 2. 4 Examples of ions ratiometric sensing platforms. (a)** *(left)*: schematic illustration of the peptide-functionalized carbon dots (f-CDs) that work as $Ca^{2+}$ nanosensors; *(middle)*: fluorescence emission spectra of f-CDs at various concentrations of $Ca^{2+}$ ($\lambda_{ex}$= 350 nm), the binding of calcium cations can quench the fluorescence emission of f-CDs; *(right)*: cell viability assay on SH-SY5Y cells incubated at various concentrations of f-CDs, for 24 and 72 hours. *Reprinted with permission from Lin et al., Sensors and Actuators B: Chemical, 2018, 273, 1654-1659. © 2018 Elsevier B.V. All rights reserved.* **(b)** *(left)*: to the right: schematic illustration of the $K^+$ nanosensors, the upconverting nanoparticles (UCNPs) are coated with PBFI loaded silica and an outer shell of $K^+$ permeable film; *(middle)*: fluorescence emission spectra of the nanosensors at various $K^+$ concentrations ($\lambda_{ex}$= 808 nm); *(right)*: CLSM micrographs of HEK 293 cells labelled with $K^+$ nanosensors, showing fluorescence emission at 400-500 nm and 500-600 nm. The potassium cation efflux, after the treatment with 5 μM nigericin, 5 μM bumetanide, and 10 μM ouabain, is verified by the fluorescence enhancement of the PBFI. *Reprinted with permission of ref. (Liu et al., Sci. Adv.6, eaax9757, 2020; figure licensed under a Creative Commons Attribution NonCommercial License 4.0 (CC BY-NC).* **(c)** *(left)*: schematic illustration of the $Cl^-$ nanosensors, CdSe/ZnS quantum dots are capped with the Chloride sensitive thiourea; *(middle):* fluorescence emission spectra of the $Cl^-$ nanosensors at increasing concentration of chloride ($\lambda_{ex}$= 425 nm); *(right):* fluorescence emission of T84 cells incubated with the nanosensors and treated with Lubiprostone, showing the efflux of chloride anions from cells. *Reproduced with permission from ref. Wang et al., Nanotechnology, 2010, 21, 055101. © IOP Publishing. All rights reserved.*

Sodium cations are the most abundant inorganic cations in human bodies. They play a fundamental role in physiology, being involved in neurotransmission, muscle contraction and blood volume regulation.[179] The $Na^+$ concentration in bodily fluids can vary a lot. The typical concentration in blood is 135-150 mM, in the interstitial fluids is around 142 mM while in the cytosol is always kept low at around 5-30 mM by the activity of ion pumps (such as $Na^+/K^+$ ATPase).[184] Fluorescent probes for $Na^+$ and $K^+$ imaging are chemically different from the $Ca^{2+}$ ones. Instead of having polycarboxylic moieties, they have a typical ether crown that is able to bind monovalent cations and a fluorescent aromatic core.[80] One of the most used molecular probe for $Na^+$ cations imaging is SBFI (Sodium-Binding Benzofuran Isophthalate), which is 20-fold more selective for $Na^+$ compared to $K^+$, and with a $K_d$ of 20 mM, making SBFI useful for intracellular imaging. Sodium Green is another $Na^+$ sensitive fluorophore, with higher selectivity over potassium (41-fold more selective) and a lower $K_d$ of 8.4 mM, that allows the detection of even smaller variation of sodium concentration. Other $Na^+$ sensitive fluorophores are CoroNa (with a $K_d$ of 80 mM), ANa1 and ANa2.[184] Dietrich et al. reported the synthesis of $Na^+$ sensitive ratiometric nanosensors based on gold nanoparticles.[185] Sodium Green was used as the sensitive dye, while Texas red was used as reference dye and both were embedded in the nanosensors by covering gold nanoparticles with layers of poly(vinylalcohol) and polyacetal. Nanosensors were incubated with CHO cells (Chinese hamster ovary cells) and $Na^+$ intracellular variations were monitored during the treatment with the ionophores nigericin, gramicidin and monesin. Wang et al. reported the synthesis of ionophore-based fluorescent ratiometric nanosensors for $Na^+$ and $K^+$, by exploiting graphene quantum dots (G-QDs).[186] G-QDs were modified with propargyl bromide and then a $Cu^{2+}$ catalysed Huisgen addition was used to coat the G-QDs with polyoxyethylene bis(azide). Sodium sensitive nanosensors were obtained by mixing PEG-GQDs with Sodium Ionophore X (NaX), sodium tetrakis-[3,5-bis(trifluoromethyl)-phenyl] borate (TFPB), bis(2-ethylhexyl) sebacate (DOS) and the Oxazinoindolines (OX-R).[187] To obtain $K^+$ nenosensors, Valinomycin was used instead of NaX. Both Nanosensors displayed a sensitivity range between 0.1 mM and 1 M. One of the main limits in the use of this chromoionophore system is the interference of pH value in the readout of the sensors. However, $Na^+$ sensors were tested on HeLa cells to evaluate the toxicity, showing good biocompatibility. To assess whether the nanosensors are suitable to detect alteration in the intracellular environment, HeLa cells were incubated with nanosensors, and treated with gramicidin[188] and carbonyl cyanide 3-chlorophenylhydrazone (CCCP),[189] which act as ionophore on cell membranes. Fluctuation in fluorescence emission was monitored by means of CLSM, showing a decrease in intracellular $Na^+$ levels.

Potassium cations ($K^+$) are the most abundant cations in the intracellular compartment, with a mean concentration of 140-150 mM, while in blood and extracellular fluids is lower at 3.5-5 mM.[190] $K^+$ cations play a crucial role in neurotransmission, muscle contraction, insulin release and are also involved in pathological events, such as epilepsy, cardiac arrhythmia and cancer.[190] There are different fluorescent molecular probes for potassium cations imaging. The most widely employed probe is Potassium-binding Benzofuran Isophthalate (PBFI), which suffers from poor selectivity between $Na^+$ and $K^+$, and an excitation peak in the far UV, with low penetration capability and potentially harmful for living cells.[191] The Asante potassium green (APG or IPG) family of fluorophores is more selective compared to PBFI, with an excitation peak in the visible light and with different $K_d$.[191] In order of overcoming the limitation of PBFI as sensitive probes, Liu et al. reported the synthesis of $K^+$ nanosensors based on upconverting nanoparticles.[192] These nanoparticles display the ability to convert two or more photons in one photon with higher energy.[193] In this case, NaYF4:Yb/Tm nanoparticles were coated with silica and then PBFI was embedded in an ion-selective polymer. This nanosensor design allowed the excitation of PBFI with near-infrared wavelength, at 800 nm, increasing the tissue penetration capability. Meanwhile, the low selectivity of PBFI was improved by the ion-selective polymer. The shielded sensors displayed a sensitivity range between 2.8 μM and 150 mM of $K^+$. Nanosensors were tested on HEK 293 cell to assess the ability to monitor fluctuations in the extracellular environment, by using nigericin and bumetanide (the fist as ionophore and the latter as inhibitor of the Na-K-Cl co-transporter).[194] Liu et al. also reported the synthesis of $K^+$ nanosensors based on silica nanoparticles.[195] The silica nanoparticles were covered with an ion-selective polymer to trap the Asante potassium green 2 (APG-2) (**Figure 2.4 b**). This nanosensor showed a sensitivity range between 1.3 μM and 150 mM and was used in both ex-vivo and in-vivo models of murine epilepsy to monitor $K^+$ variations. Ruckh et al. reported the synthesis of a ratiometric $K^+$ nanosensor based on QDs.[196] The system was composed by two QDs species with non-overlapping emission spectra, a non-fluorescent chromoionophore and an ionophore. The protonation state of the chromoionophore affected its absorption spectra, with a consequent effect on the QDs emission spectra. The final readout was derived from the ratio of the two QDs species. The nanoparticles based on this system and embedded with a plasticizer displayed a sensitivity range between 2 and 120 mM. The nanoprobes were tested on HEK 293 cells for assessing their ability to monitor $K^+$ fluctuations in the extracellular environment.

The chloride anion ($Cl^-$) is the most abundant anion in human bodies and the most important in electrophysiological regulation. It plays a crucial role in neurotransmission and is also involved in pathological conditions, such as cystic fibrosis.[80] The chloride concentration inside cytosol can vary a lot in different cell lines, while in plasma and in interstitial the concentration is kept constant by kidneys filtration at approximately 100 mM.[197] Most of the electrophysiological studies conducted on the Cl$^-$ role in physio-pathological processes are performed with patch clamp, ion-selective electrodes and chloride radioisotope. However, there are different fluorescent molecular probes, commercially available, for Cl$^-$, such as lucigenin (*bis*-N-Methylacridinium nitrate), SPQ (6-Methoxy-N-(3-sulfopropyl)quinolinium), MEQ (6-methoxy-N-ethylquinolinium), and BAC (10,10' Bis[3-carboxypropyl]-9,9'-biacridinium dinitrate). Ruedas-Rama *et al*. reported the synthesis of Cl$^-$ nanosensors based on semiconductor QDs and lucigenin.[198] Hexadecylamine-capped CdSe/ZnS QDs were modified with 3-mercaptopropionic acid (MPA), and lucigenin was bound to nanoparticles by simple electrostatic interactions, between the positively charge acridine and negatively charged MPA. The lucigenin fluorescence was quenched by Cl$^-$ as a result of a charge transfer mechanism. However, the QDs-Lucigenin conjugate showed increasing fluorescence emission in the QDs spectrum, related to the competing action of FRET between QDs and lucigenin, and charge transfer between lucigenin and Cl$^-$. The calibration showed how the nanosensors can be used by means of FLIM or with the ratio between QDs emission and lucigenin emission. The system proved to be selective toward other anions and sensitive in the Cl$^-$ concentration range of 0.5 mM to 50 mM. Wang *et al*. reported the synthesis of Cl$^-$ nanosensors based on semiconductor quantum dots and a novel thiourea moiety. The 1-(2-mercapto-ethyl)-3-phenyl-thiourea was synthesized and used for the capping of CdSe/ZnS QDs (**Figure 2.4 c**).[199] The nanosensors displayed sensitivity in the range of Cl$^-$ concentration of 2 mM to 130 mM. A comparison between MEQ and nanosensors was performed on CF-PAC human epithelial cells. The two probes were embedded into liposomes and administered to CF-PAC cells treated with glibenclamide (as a chloride channels inhibitor). The nanosensors exhibited better sensitivity in monitoring intracellular Cl$^-$ fluctuations, compared to MEQ, employing FLIM analyses.

## 2.5 Biomarkers

Macromolecules such as nucleic acids, proteins, metabolites, isoenzymes and hormones are well recognized as characteristic signature of cancer onset and progression.[200] Biomarkers are molecules or substances that can indicate the presence or progression of a disease, and are classified into three main categories in clinical practice: **(i)** diagnostic biomarkers, which are used for disease detection; **(ii)** prognostic biomarkers, which provide information about the likelihood of disease recurrence; and **(iii)** predictive biomarkers, which can help determine the patient's response to cancer treatment.[200] The most used methods to detect and quantify biomarkers are based on enzyme-linked immunosorbent assay (ELISA)[201] and polymerases-chain-reaction (PCR)-based protocols.[202] While widely accepted as crucial procedures in cancer diagnosis and treatment, these technologies are not without limitations. One significant challenge is the slow reaction mechanism of detection, which can delay diagnosis and treatment initiation. Additionally, the exorbitant cost of reagents required for these techniques can result in high expenses for patients, limiting their accessibility and affordability.[201,202] Currently, there is only a limited number of published studies on ratiometric fluorescence-based nano-systems for cancer biomarkers.[203]

A strong correlation between extracellular pH acidification and the increased expression of tumor-related proteases has been associated with the invasion and dissemination of cancer cells in other organs.[204,205] The quantification of matrix metalloproteases-2 (MMP-2) in blood is still a great challenge because of the complexity of the biological fluid. In this context, Wang *et al*.[206] engineered an upconversion FRET-based biosensor to specifically target MMP-2. The authors recorded a linear and proportional relationship between MMP-2 concentration and fluorescence recovery of the sensing system in the range 10–500 pg/mL. The validation of the bioanalytical sensors was carried out by collecting human plasma and whole blood samples. In the last twenty years, several efforts have been engaged to construct sensing platforms capable of tracking the activity of proteases and exploiting their detection for the early diagnosis of cancer. An example is represented by the FL–gold nanoparticle activatable probes.[207–209] In 2008 and 2009 Lee and collaborators developed a NIR-FL gold nanoparticle, which showed to be selective for matrix metalloproteases-2 (MMP-2). Thanks to its specificity, the developed probe allowed the simple monitoring of the activities of MMP-2, both *in vitro*, using HepG2 cell line, and *in vivo*, adopting mice bearing SCC7 (squamous cell carcinoma) tumors.[210,211] In another work by Yin *et al*.[212], a MMP-2 activatable probe was prepared by covalently coupling a near infrared dye (Cy5), a quencher (QSY21), a tumor targeting peptide (Cyclic Arg-Gly-Asp) and a radionucleotide $^{125}$I-labeled peptide substrate. The developed probe, which light-up upon proteolytic cleavage operated by active MMP-2, was employed for the accurate detection, via NIRF and single-photon emission computed tomography (SPECT) imaging techniques, of the metastatic lymph nodes (MLNs) in mice bearing murine breast carcinoma cell line 4T1, before and during treatment with a MMP-2 inhibitor. The above reported examples represent a fundamental application of highly selective and sensitive FL-based platforms and represent milestones in the field of ratiometric probes. In 2018, Ma *et al*. created an FL ratiometric probe to track matrix metallopeptidase-9 (MMP-9) activity and extracellular pH, both *in vitro* and *in vivo* (**Figure 2.5 a**).[213] The sensing platform was architected as follows: biocompatible PEGylated iron oxide ($Fe_3O_4$) magnetic NPs were chosen as sensor support material; the sensing units, represented by the pH-sensitive naphthalimide dye ANNA ($\lambda_{ex}$= 455 nm; $\lambda_{em}$= 510 nm), labelled with a peptide substrate of MMP-9 (GGKGPLGLPG), and the reference dye Cy5.5 ($\lambda_{ex}$= 675 nm; $\lambda_{em}$= 695 nm) were covalently linked to PEGylated $Fe_3O_4$ NPs. The FRET-based mechanism determined the FL "off" state when the pH fluorophore ANNA was bound onto the surface of $Fe_3O_4$ NPs. The cleavage of the peptide substrate, operated by the MMP-9, contributed to the FL enhancement of ANNA dye, which turned in the "on" state because of its protonation. Employing time-lapse CLSM, the ratiometric linear regression obtained by plotting $I_{ANNA}/I_{Cy5.5}$, allowed the authors the detection and quantification of the MMP-9 following the incubation of the ratiometric sensing probe in human colorectal cancer cell line LS180, which is known to overexpress MMP-9. To further validate the analytical platforms, the authors injected the ratiometric probe via the rat vein of tumor-bearing mice. The results showed that the nanoprobes were activated in the tumor site 4 hours post-injections and the pH mapping and MMP-9 quantification, executed by images analysis, confirmed the possibility of the real-time monitoring of multiple TME targets.

Another interesting target in clinical routine is telomerase, a transcriptase responsible for unlimited cancer proliferation.[214,215] For this reason, today it is considered a diagnostic and prognostic biomarker. A QDs-based ratiometric FL sensor, with FRET mechanism, was developed by Ma *et al.* to specifically target intracellular telomerase activity (**Figure 2.5 b**).[216] The nanoprobe was constituted of a core-shell streptavidin-modified cadmium selenide/zinc sulphide core-shell (CdSe/ZnS) QDs (QDs$_{SA}$) ($\lambda_{ex}$= 235 nm; $\lambda_{em}$= 600 nm), functionalized with a telomerase primers (TP) and a signal switching sequence (SS), which was designed to form a hairpin filament complementary to telomerase. To obtain a sensing system, the SS was labelled at its 5' end with cyanine 5 (Cy5) ($\lambda_{ex}$= 630 nm; $\lambda_{em}$= 665 nm) that acted as an FL acceptor. Thus, once assembled, the nanoprobe exhibited only the FL of Cy5 because of the FRET-phenomenon; in contrast, in presence of telomerase, the TP was recognized and elongated by telomerase itself, hybridized with SS to form a double-strain and disassembled from the nanoprobe providing the switch-on of QDs$_{SA}$ and amplification of the FL signal, yielding in a significant ratiometric FL change ($F_{QDsSA}/F_{Cy5}$). After the optimization of the best sensing conditions, the nanoprobes were incubated with different interfering biomolecules such as bovine serum albumin (BSA), immunoglobulin G (IgG), lysozyme, thrombin, trypsin, ATP, RNA, and Bst DNA polymerases, and for no one of them the probe showed significant FL changes, thus presenting a strong selectivity for telomerases. To map the telomerase activity *in vitro*, the authors employed two types of cells: cervical carcinoma HeLa cells and human hepatocyte cells (L-O2). As expected, in the cancerous HeLa cells, the green fluorescence of QDs$_{SA}$ was visible, reflecting a high telomerase activity, while the opposite result was obtained in healthy L-O2 cells. These results confirmed that the intracellular telomerase can now be detected and monitored in a spatio-temporal manner thanks to ratiometric FL platforms to be used, in the near future, as promising tools for diagnostic cancer screening and telomerase-targeted anticancer drugs.

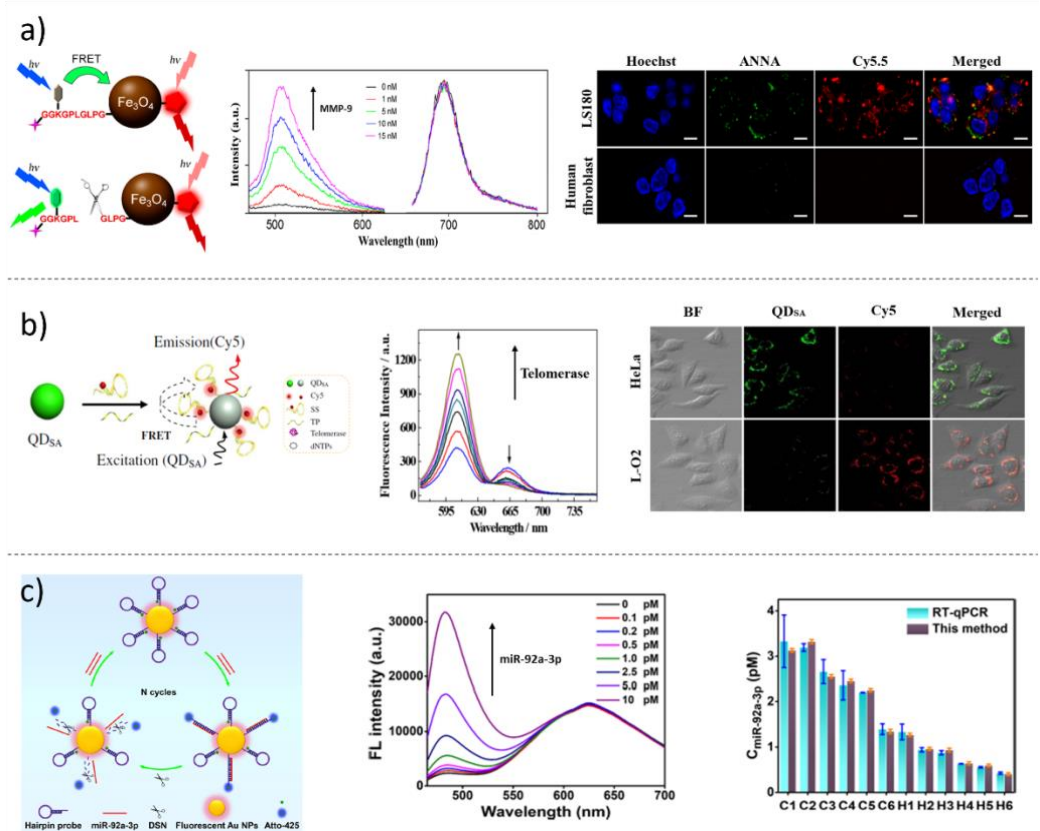

**Fig. 2. 5 Examples of biomarkers sensing nano-platforms. a)** *(left):* schematic illustration of the working mechanism of Fe$_3$O$_4$ MMP-9 activity sensing NPs; *(middle)*: fluorescence spectra recorded after the nanoprobes were incubated with different concentrations of activated MMP-9; *(right)*: confocal microscopy images of LS180 cells (top row) and human fibroblast control cells (bottom row) obtained after incubation with the nanoprobe for 6 h and then imaged through different channels according to the dye emissions (cell nuclei were stained with Hoechst, and the scale bar corresponds to 10 μm). *Reprinted with permission from Ma et al., J. Am. Chem. Soc. 2018, 140, 1, 211–218. Copyright © 2018, American Chemical Society.* **b)** *(left)*: the QDs$_{SA}$@DNA nanoprobe for monitoring of telomerase activity *in situ*; *(middle)*: fluorescence emission spectra of the designed QDs$_{SA}$@DNA nanoprobe (100 nM) in response to telomerase extraction from different numbers of HeLa cells; *(right)*: confocal fluorescence microscopy imaging of HeLa and L-O2 cells incubated with the QDs$_{SA}$@DNA nanoprobe for 4 h. The concentration of the added QDs$_{SA}$@DNA nanoprobe was 100 nM. Scale bar: 25 μm. *Reprinted with permission from Ma et al., Anal Bioanal Chem 414, 1891–1898 (2022). Copyright © 2022, Springer-Verlag GmbH Germany, part of Springer Nature.* **c)** *(left)*: schematic of the ratiometric fluorescent detection of miR-92a-3p based on fluorescent Au-NPs and DSN-assisted signal amplification; *(middle)*: fluorescence spectra of the biosensor under different concentrations of miR-92a-3p; *(right)* comparison of the exosomal miR-92a-3p concentrations of CRC patients and healthy controls detected by RT-qPCR and this method (n = 3, mean ± s.d.). C1–C6 represent CRC patients; H1–H6 represent healthy controls. *Reprinted with permission from Sun et. al., Bioconjugate Chem. 2022, 33, 9, 1698–1706. Copyright © 2022, American Chemical Society.*

With the advent/irruption of cancer liquid biopsy, the possibility of screening and detecting a high number of circulating tumor biomarkers, such as extracellular vesicles, proteins, nucleic acids, and microRNAs (miRNA), by using simply a blood sample, has become true.[217,218] A smart example of sensing colorectal cancer (CRC)-associated exosomal miRNA (miR-92a-3p), by developing ratiometric FL nanoparticles, was described in the work of Sun and collaborators (**Figure 2.5 c**).[219] The nanoprobe was composed of a hairpin DNA, labelled with sulfhydryl Atto-425 ($\lambda_{ex}$= 443 nm; $\lambda_{em}$= 482 nm) on the 5' end and 3' end, and conjugated to the surface of a fluorescent gold NPs (Au-NPs) ($\lambda_{ex}$= 320 nm; $\lambda_{em}$= 625 nm). In this way, the FL emissive peak of Atto-425 was quenched, via a FRET mechanism, by the FL of the Au-NPs. In contrast, in presence of miR-92a-3p, Atto-425 was unbound from the surface of the Au-NPs, recovering its fluorescence. The detachment of Atto-425 was coordinated by the activity of a duplex-specific nucleases (DSN), which cleaved the DNA filament in a miR-92a-3p/DNA heteroduplex, keeping the RNA fragment intact. This process drove to a signal amplification because the free miR-92a-3p could open another hairpin DNA, repeating in this way the sensing cycle. The ratiometric calibration curve, carried out plotting the $I_{Atto425}/I_{Au-NPs}$, in a concentration range of 0.1-10 pM miR-92a-3p, presented a linear regression fit of 0.995 and a limit of detection of 45 fM, thus improving the sensitivity in miRNA detection. Exosomal miR-92a-3p were extracted from 3 CRC patients' sera and 6 healthy controls' sera clinical samples, after lysis of exosomes. The results in detection and quantification of miR-92a-3p extracted from patients' and controls' samples using the ratiometric FL nanosensors were consistent with those obtained adopting real time- quantitative PCR (RT-qPCT). Thus, the biotechnological analytical platform can be considered a promising device and adopted as a potential tool in clinical diagnosis.

# 3. Hybrid materials/systems including fluorescent nano-microparticles for biomedical applications

In these days and age, monitoring the metabolic variations in TME *in toto* and providing, at the same time, a real-time detection at a single-cell scale are still challenging cues not only in cancer, but also in tissue engineering and regenerative medicine.[220] The combination of diverse nanotechnologies has recently led to the development of integrated sensing devices able to reveal the spatio-temporal behaviour of cells using high-resolution and computational methods. In this chapter, particle-reinforced biocomposites are treated in which the dispersed phase is represented by smart fluorescent nano- and microparticles for medical therapeutic and/or diagnostic purposes. The manipulation of parameters such as the volume ratios of the components, matrix type, included particle size and nature, geometry, orientation and distribution offer a wide design flexibility. Matrix materials of biomedical interest are represented by fibrous matrices based on biopolymers, since they replicate the organization and biological behaviour of the extracellular matrix.[221] ECM-like fiber mats can be produced by electrospinning, which is a cost-effective method to fabricate fibers having diameters ranging from nanometers to microns.[222] Electrospun fibers exhibit many advantages including tunable composition and size, tunable alignment, and the possibility of being loaded with drugs and stimuli-responsive nanomaterials for enabling controlled and sustained release via physiological or physical stimuli.[223] Because of their size and surface features, nanofibers show high surface-to-volume ratio and porosity, which favour the transport of small molecules as ions, making them a particularly attractive platform for the development of ultrasensitive sensors. Indeed, nano- and microparticle-based sensors can be dispersed in the polymer solution and entrapped by electrified jets within the lumen of the electrospun fibers forming functional optical regions within the mats. The use of electrospinning to produce sensing matrices has been successfully reported by many groups.[224,225] For example, pH-sensing electrospun fiber scaffolds were produced by embedding pH sensors for the ratiometric measurement of local proton concentration, with high spatial resolution and in a fast and non-invasive manner.[226,227] The fluorescence changes of the functional regions were correlated with $H^+$ concentration during spatio-temporal measurements of extracellular acidity of pancreatic tumor and stromal co-cultures. Single-cell fermentation flux analysis via constraint-based inverse modelling evidenced that $H^+$ trafficking was strongly heterogeneous with just few cells showing high activity, and therefore, responsible for a large fraction of the pH gradient in the cell culture (**Figure 3a**).[227] Nano- and microparticles have been also incorporated in biopolymer-based hydrogels, which is a widely used class of materials in tissue engineering, ophthalmic, wound healing and drug delivery thanks to properties as biocompatibility and biodegradability.[228] Particle sensors can be easily dispersed within hydrogels, and because of their macromolecular polymer network structure and hydration, the mobility of small ions does not decrease significantly compared to their diffusion in aqueous solutions.[229,230] Among hydrogels, alginate is a naturally occurring biopolymer that founds many applications owing to its biocompatibility, low cost, ease of gelation and optical transparency that makes it highly suitable even for microscopy applications. Very recently, alginate-based three-dimensional microgels were produced via electrostatic droplet encapsulation method to embed FITC-RBITC SiO$_2$ pH sensors together with pancreatic tumor and/or stromal cells (**Figure 3b**).[231] The method involved the use of high voltage to obtain droplets of average diameter of around 200 μm, which were crosslinked in a solution of calcium chloride. Extracellular pH metabolic variations were monitored by means of 4D (x, y, z, t) live-cell imaging showing that pH was cell line-specific and time-dependent. In addition, differences in acidification were measured in 3D mono- against 3D cell co-cultures, suggesting the existence of a cancer-stromal cells crosstalk, resulting from metabolic cell reprogramming towards a glycolytic phenotype.[232] The same droplet encapsulation method with alginate has also been used to incorporate pH-sensitive carbon nanoparticles for measuring pH during bacterial cultures. The ratiometric detection was assessed by calculating the ratio 550 nm/450 nm of fluorescence emissions that was plotted against incubation time to obtain the growth rate of bacteria. Results showed that the emission ratio increased, and therefore pH decreased over time, reflecting the growth of bacteria.[233] Luminescent amphiphilic carbon nanoparticles (CDs) were also embedded within an ascorbic acid derivative hydrogel for the detection of reactive oxygen species (ROS). ROS induced oxidation of the ascorbic acid units with consequent collapse of the hydrogel, aggregation of the CDs and therefore quenching of their luminescence, monitorable under ultraviolet (UV) excitation. CDs–hydrogel was applied as sensing platform to detect *in vitro* the presence of ROS from HeLa cells exposed or not to 5-fluorouracil (5-FU). 5-FU is a chemotherapeutic known for generating intracellular ROS, leading to apoptosis of cancer cells.[234] Therefore, quenching of the CDs-hydrogel revealed the efficacy of the treatment, evidencing the potential of the system for drug screening applications.[234] Recently, Delic et al. developed another fluorescent composite material from CDs. To preserve the intrinsic fluorescence in the dried solid state and in aqueous solution at a broad pH range (pH=3-12), nanoparticles were dispersed and embedded throughout silica particles, obtaining, thanks to hydrothermal treatment at low temperature in presence of urea and citric acid, a final fluorescent macroporous hollow structure, ideal for drug storage and delivery systems.[235] Another ratiometric fluorescent microgel was produced by Li et al. from polyurethane (PU), a widely used material in medicine for its stability and biocompatibility. pH-sensing nanoparticles were loaded within PU obtaining a final particle size distribution of around 75 μm and spherical morphology. Nanoparticles were synthesized by cross-linking of denatured pH-sensitive bovine serum albumin proteins and pH-insensitive Nile Red as a reference and exhibited linear reversible fluorescence response to pH between 6 and 10. The system was used to study, through colorimetric maps, local extracellular pH during biomaterial degradation since this phenomenon could have a great influence on the surrounding cells, for example it could affect the balance between bone formation and resorption.[236] Fluorescent microparticles were also included in a polymer layer to create an integrated sensor for monitoring oxygen gradients of hypoxic tumors. Specifically, silica microparticles were first absorbed with oxygen-sensitive (Ru(dpp)$_3$Cl$_2$) and insensitive (Nile Blue chloride) fluorophores, and then mixed in polydimethylsiloxane (PDMS). The mixture was then poured onto the pillars of a microfluidic device recreating insulated oxygen conditions on the two sides of a monolayer of human breast cancer cells MCF7 and therefore hypoxic levels in the monolayer. After 24 hours of cell culture, an enhanced fluorescence of ruthenium was observed at the center of the pillar, which was decreasing radially. The ratiometric ruthenium by nile blue intensity was plotted against the radial distance and molecules and proteins regulated by hypoxia were immunostained and correlated with the oxygen gradient.[237] To monitor oxygen concentration within large 3D scaffolds Wilson et al. reported the synthesis via organic-in-oil suspension of fluorescent hydrogel MPs containing palladium(II) meso(tetracarboxyphenyl)-porphine as oxygen-sensitive fluorophore and Alexa Fluor 633 carboxylic acid tris(triethylammonium) salt as reference fluorophore. MPs were encapsulated into a cellularized hydrogel scaffold and oxygen gradients were measured via ratiometric imaging, showing that the signals were robustly photostable and unaffected by hydrogel thicknesses of over 2 mm.[238] Another approach for monitoring oxygen gradients within a hydrogel environment providing real-time information for ensuring efficient cell functions involved the development of functional fluorescence-based nano-oxygen particles (FNOPs). Specifically, Pluronic F127-grafted polystyrene beads (PSBs) were linked with the commercially available oxygen-sensitive fluorescent molecule Ru(dpp)$_3$Cl$_2$ through hydrophobic interaction. FNOPs were applied for measuring oxygen concentration of RIN-m5F/HeLa cell lines in hydrogel spheres of 700–1000 μm in diameter generated by the electrospray technique for more than 5 days.[238] Oxygen concentration can also be probed by using a confocal phosphorescence life-time microscope (PLIM).[98] Commercially available oxygen microsensor beads were mixed with the cell suspension, with or without the presence of gelatin methacryloyl (GelMA) as photocrosslinkable hydrogel matrix, prior to seeding in static 2D, 3D cultures or in single or 8-channel array perfusion chips. Oxygen-dependent phosphorescence decay profiles of the microsensors can be converted, through a calibration curve, to oxygen concentration to map and predict oxygen distributions for different cell densities, media and cultured cells (**Figure 3c**).[239] Finally, hybrid systems including pH-fluorescent sensors were developed also by combining different technologies, in order to probe 3D cell growth and tissue regeneration.[240] Capsules based on SNARF-1 were included in a 3D additive manufactured scaffold with controlled geometry and porosity providing a real-time detection of the acidification of human mesenchymal stromal cells. pH in the cell microenvironment showed a reduction after 7 days, which was more prominent at the edges of the 3D scaffolds evidencing the importance of considering the position of the sensors within a scaffold to detect smaller pH gradients that exist spatially around the cells.[241]

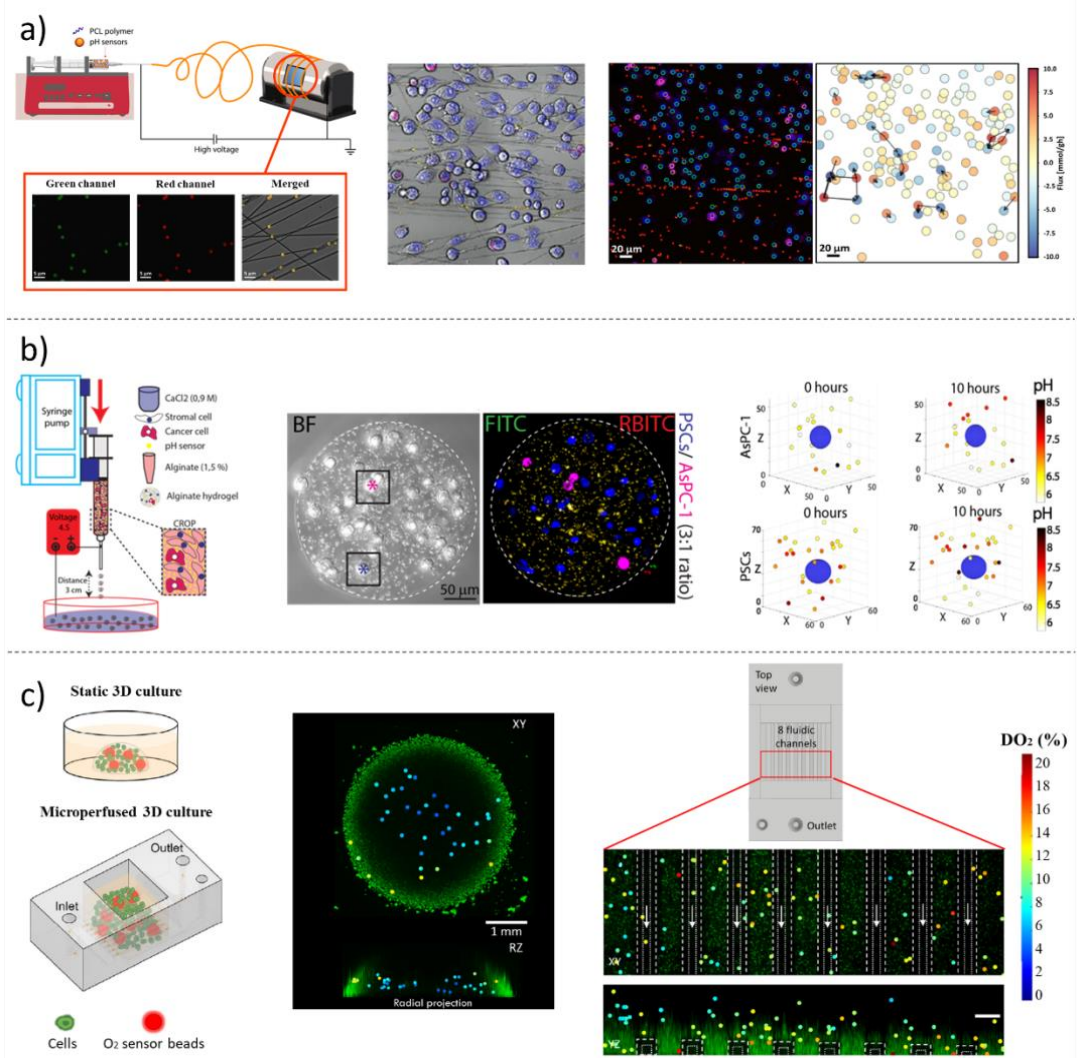

**Fig. 3 Examples of hybrid materials/systems including fluorescent nano-microparticles for biomedical applications. a)** *(left)*: Sketch showing the fabrication of electrospun polycaprolactone (PCL) fibers embedding ratiometric $SiO_2$-based microparticle sensors and representative CLSM micrographs showing PCL nanofibers embedding pH sensors (deposition time = 30 s). FITC (green channel), RBITC (red channel), and overlay with bright-field (BF, gray channel) are shown. Scale bar: 5µm; *(middle):* representative CLSM image showing cells co-cultured on pH-sensing fibers and analyzed by CLSM time-lapse imaging (x, y, z, t; t = 6 h) (nuclei are shown in blue, and cell membranes are shown in magenta for tumor cells). *(right):* results of the segmentation showing the detection of the single pH sensors (red circles), AsPC-1 cells (green circles), and CAF cells (yellow circles) and corresponding reconstruction of the cell fluxes through physically constrained statistical inference, with a relative colormap. Scale bar: 20 µm. *Reprinted with permission from Onesto et al., ACS Nano 2023, 17, 3313–3323; figure licensed under CC-BY 4.0 https://creativecommons.org/licenses/by/4.0.* **b)** *(left)*: schematic illustration of the microencapsulation system for the generation of 3D spherical hydrogels embedding pH sensors, tumor and stromal cells; *(middle):* maximum intensity projection of 3D time-lapse CLSM acquisitions of alginate hydrogel including FITC/RBITC pH-sensors (yellow), tumor cells (magenta) and stromal cells (blue). Bright-field (BF, grey). Scale bar: 50 µm. *(right):* 3D scatter plots of the pH-sensors around a selected tumor and cancer cells at time 0h and 10h, with relative pH colour-maps. *Reprinted with permission from ref. (Rizzo et al., Biosensors and Bioelectronics, 2022, 212, 114401; figure licensed under CC-BY 4.0 https://creativecommons.org/licenses/by/4.0. Copyright © 2022 The Authors. Published by Elsevier B.V.* **c)** *(left)*: the cell mixture is dispensed into the targeted 3D static culture, and 3D dynamic culture within a chip supporting microperfusion (perfusion pathway illustrated by arrows); *(middle):* CLSM PLIM imaging of the embedded oxygen sensor beads and subsequent conversion to the corresponding local oxygen concentration is shown as colour coded concentrations overlaid with green fluorescence from calcein AM staining of metabolically active cells in lateral (XY) and radial (RZ) projections. Hep G2 cells at $20 \times 10^6$ cells per mL embedded in a hydrogel of 7.5% w/v GelMA in medium; *(right):* oxygenation in a 3D tissue model with an array of 8 perfused microfluidic channels (inner dimensions $140 \times 140$ µm$^2$) and projected bottom view of sensor beads at all elevations overlaid on a projected confocal fluorescence micrograph of live-stained cells. *Reproduced from Wesseler et al., Lab Chip, 2022, 22,4167 with permission from the Royal Society of Chemistry.*

## 4. Challenges and future perspectives of ratiometric FL sensors in oncology

The feasibility and versatility of accurate and sensitive ratiometric FL sensing systems have been explored in this review with a particular focus on their application for in vitro and in vivo spatiotemporal mapping of TME parameters, highlighting their potential for application in cancer diagnostics and therapeutics. Among many milestones achieved in this era, the possibility of adopting FL ratiometric sensors in clinical practice has become a reality and is now of great impact in surgery.[242] In fact, the intraoperative FL-guided surgery is today capable to discriminate a healthy tissue from a cancerous one, resulting in a more accurate resection of the diseased area.[243–246] The main advantage coming from such fine surgical resection is the complete eradication of the tumor. This procedure is therefore strictly linked to the reduction of tumor metastasis circulation, from a primary lesion towards other organs. In addition to this several other reports in literature exhibit the use of ratiometric FL probes for mapping tumor's margins in biopsy tissues[247] or during surgery.[248] Frequently, surgery is accomplished by pharmacological therapies, therefore another benefit that could be added to the clinical routine would regard the use of ratiometric FL sensing systems capable of locally monitoring drug release.[249] An example is represented by a recently developed ratiometric drug delivery system comprising a target-specific antibody for selective delivery to cancerous cells linked to a "drug–switchable dye" conjugate and a reference dye for ratiometric fluorescence monitoring of drug release.[250–253] By using this system the authors showed intensity-based monitoring of drug distribution and accumulation in vitro and in vivo as well as ratiometric measurements of drug release in vitro. Possible future developments could also include the fabrication and application of barcode ratiometric sensors for in vitro/in vivo multiplex spatiotemporal analysis of key metabolic TME parameters (e.g., pH and oxygen) or metal ion concentration (e.g., calcium). A multiplexed non-invasive ratiometric sensing would allow for continuous and online recording of multiple cellular input signals influencing the physiological and pathological states of the body, thus improving our understanding of cell–cell interactions and cell's response to therapies.

## Conclusions

Fluorescence-based sensing and imaging have emerged as cutting-edge technology for ratiometric fluorescence devices, providing accurate *in vitro* and *in vivo* detection and monitoring of TME parameters and related cancer biomarkers. The ability to capture cell-cell interactions and single-cell behaviour in different time points, qualitatively and quantitatively, allows for a deeper understanding of the complex world of cancer. This approach provides researchers and clinicians with precise and informative tools to fight against this widespread disease.

The design of smart ratiometric FL probes, in the form of nano- and microparticles, may seem straightforward at first glance. However, the working principles and effectiveness of a developed sensor platform depend on overcoming various obstacles that must be considered when imaging is the ultimate goal. Some of them are photostability, leaching, and autofluorescence generated by macromolecules in living organisms, as well as perturbations caused by operator, instrumental and environmental conditions, which can occur inevitably during the whole experiments. The strength of the ratiometric FL method stands in its self-calibration, achieved by employing an indicator signal and a reference signal, or two reversible signals, knocking down errors and enhancing accuracy and precision. In addition, the ongoing research and development of novel sensing probes spanning a wide range of options including organic dyes, quantum dots, nanoclusters, and metal-organic frameworks, paired with diverse synthetic methodologies such as encapsulation and layer-by-layer (LbL) deposition of sensing units, the ability to customize particles with biocompatible molecules, and the ability to tune their size, shape, charge, and matrix support, have made ratiometric FL tools versatile platforms for sensing the TME both *in vitro* and *in vivo*.

Today, the main players of the TME, including both cellular and non-cellular components, have been widely studied and even though their leading functions in cancerogenesis and progression have been interpreted, the entire tumor pool still remains a constant-evolving topic. Phenomena such as acidosis, hypoxia, ROS generation, and ions fluxes variations are known to be strictly interconnected with each other and all of them represent targets that are selectively hit by therapeutic protocols. Meanwhile, in the literature it is possible to find many ratiometric FL tools for the detection of intracellular and extracellular pH of cancer cells, more forces need to be engaged to develop ratiometric FL sensors for monitoring and real-time measurements of tumor hypoxia. On the other hand, intrinsically fluorescent nanomaterials, such as semiconductor quantum dots and metal nanoclusters, have emerged as promising candidates for the design of highly sensitive and photostable ratiometric FL systems. These nanomaterials have been successfully employed to monitor the physio-pathological changes in pH, dissolved oxygen, ROS, and ion levels in tumor cells, and have greatly assisted in the development of analytical platforms that are being used as point-of-care devices for the screening and targeting of diagnostic, prognostic, and predictive biomarkers. To overcome the challenge of mimicking the complex TME and capturing its metabolic dynamics in a spatiotemporal manner, ratiometric FL sensors have been successfully integrated into various biocompatible and easily-handled polymer matrices to create 2D or 3D sensing platforms. This innovative approach offers the potential for high resolution and reliable real-time measurements of cell metabolism during drug treatment, allowing for a more comprehensive understanding of the TME and cell metabolic dynamics.

The future prospective is called "precision medicine": thanks to the irruption of the nanotechnologies, which in turn are running, day by day, towards more sophisticated and ultrasensitive platforms for boosting and gaining finest and trustworthy quantification of the TME in its complexity, accomplished with its related molecular biomarkers, the cancer treatments will soon progress towards targeted therapies, the new revolutionary frontier.

## Conflicts of interest

There are no conflicts to declare.

## Acknowledgements

The authors acknowledge the funding from the European Research Council (ERC) under the European Union's Horizon 2020 research and innovation program ERC Starting Grant "INTERCELLMED" (contract number 759959), the European Union's Horizon 2020 research and innovation programme under grant agreement No. 953121 (FLAMIN-GO), the Associazione Italiana per la Ricerca contro il Cancro (AIRC) (MFAG-2019, contract number 22902), the "Tecnopolo per la medicina di precisione" (TecnoMed Puglia) - Regione Puglia: DGR n.2117 of 21/11/2018, CUP: B84I18000540002). The Italian Ministry of Research (MUR) under the complementary actions to the NRRP (PNC0000007) "Fit4MedRob- Fit for Medical Robotics" Grant (contract number CUP B53C22006960001) and the MUR NRRP "National Center for Gene Therapy and Drugs based on RNA Technology" (Project no. CN00000041 CN3 RNA). AC also acknowledges the Jain University, Bangalore (India) for its support through the provision of seed fund assistance.